\documentclass[twocolumn]{aastex631}

\usepackage{enumitem}
\usepackage{mathtools}

\def \mathbi#1{\textbf{\em #1}}

\shorttitle{Low-viscosity warped discs around black holes}
\shortauthors{Drewes \& Nixon}

\begin{document}

\title{On the dynamics of low-viscosity warped discs around black holes}

%\correspondingauthor{C.~J.~Nixon}
%\email{cjn@leicester.ac.uk}

\author{N.~C.~Drewes}
\affiliation{Department of Physics and Astronomy, University of Leicester, Leicester, LE1 7RH, UK}

\author[0000-0002-2137-4146]{C.~J.~Nixon}
\affiliation{Department of Physics and Astronomy, University of Leicester, Leicester, LE1 7RH, UK}

\begin{abstract}
Accretion discs around black holes can become warped by Lense-Thirring precession. When the disc viscosity is sufficiently small, such that the warp propagates as a wave, then steady-state solutions to the linearised fluid equations exhibit an oscillatory radial profile of the disc tilt angle. Here we show, for the first time, that these solutions are in good agreement with three-dimensional hydrodynamical simulations, in which the viscosity is isotropic and measured to be small compared to the disc angular semi-thickness, and in the case that the disc tilt---and thus the warp amplitude---remains small. We show using both the linearised fluid equations and hydrodynamical simulations that the inner disc tilt can be more than several times larger than the original disc tilt, and we provide physical reasoning for this effect. We explore the transition in disc behaviour as the misalignment angle is increased, finding increased dissipation associated with regions of strong warping. For large enough misalignments the disc becomes unstable to disc tearing and breaks into discrete planes. For the simulations we present here, we show that the total (physical and numerical) viscosity at the time the disc breaks is small enough that the disc tearing occurs in the wave-like regime, substantiating that disc tearing is possible in this region of parameter space. Our simulations demonstrate that high spatial resolution, and thus low numerical viscosity, is required to accurately model the warp dynamics in this regime. Finally, we discuss the observational implications of our results.  
\end{abstract}


\section{Introduction}

Accretion on to black holes is a major theme in modern astronomy. The black holes may be of stellar masses, typically occurring in binary systems with the accretion supply coming from a companion star through winds or Roche lobe overflow. Alternatively the black holes may be supermassive and located in galaxy centres with a plentiful supply of gas and stars as fuel. Accretion on to supermassive black holes powers the most luminous continually emitting objects in the universe, active galactic nuclei (AGNs). In both cases, the energy and momentum output of the accretion process has a marked effect on their surroundings including contributing to the epoch of reionization, affecting star formation rates and galaxy evolution, and driving metal enrichment of the intergalactic medium.

For these reasons, and others, significant effort is continually expended exploring the dynamics and evolution of accretion flows on to black holes. In this work we focus our attention on the specific case of accretion on to a spinning (Kerr) black hole who's rotation axis is misaligned with respect to the rotation of the accretion disc. In this case, the disc orbits experience Lense-Thirring precession, a relativistic frame dragging effect. \cite{Bardeen:1975aa} investigated this effect on an accretion disc around a spinning black hole, and found that the disc achieves a warped shape, with the inner disc aligned to the black hole spin and the outer disc remaining in the original misaligned plane.

There are two distinct ways in which a warp can be communicated through the disc: through diffusion or wave-like propagation \citep{Papaloizou:1983aa}. The diffusive case occurs when the dimensionless disc viscosity parameter \citep{Shakura:1973aa} is larger than the angular semi-thickness of the disc ($\alpha > H/R$). When the warp diffuses through the disc, the disc typically warps such that the inner parts align with the black hole, while the outer disc retains its initial tilt; this disc shape is known as the Bardeen-Petterson effect. In recent years a good understanding of the dynamics of warped discs in the diffusive regime has been obtained by detailed analytical models \citep[e.g.][]{Ogilvie:1999aa,Ogilvie:2000aa,Ogilvie:2013aa,Dogan:2018aa} and numerical simulations \citep[see, for example,][and references therein]{Lodato:2010aa,Raj:2021aa}. On the other hand, when the wave-like propagation of the warp is not suppressed by the viscosity in the disc ($\alpha < H/R$), discs around black holes can establish a different steady-state shape in which the tilt angle of the local orbital plane exhibits an oscillatory profile with radius \citep{Ivanov:1997aa}. The dynamics of warped discs in the low-viscosity case is less certain as there is no complete nonlinear analytical theory \citep{Ogilvie:2006aa}, and the analytical work is typically performed with linearised equations of motion \citep[e.g.][]{Papaloizou:1995aa,Ivanov:1997aa,Lubow:2000aa} with some authors focussing on specific nonlinear effects \citep[e.g.][]{Gammie:2000aa} which have been followed up with targetted numerical simulations \citep[e.g.][]{Paardekooper:2019aa,Deng:2021aa}.

Here we focus on the low-viscosity case. Specifically, here we are concerned with the wave-like case of warp propagation and the radial tilt profile that is expected to occur for such discs around a spinning black hole. For low-viscosity discs with $\alpha < H/R$, \cite{Papaloizou:1995aa} showed that warp waves propagate through the disc with a wave speed approximately half the sound speed. \cite{Ivanov:1997aa} analyse the stationary shape of a misaligned disc around a spinning black hole and find an oscillatory radial profile for the disc tilt. \cite{Demianski:1997aa} find similar results using a time-dependent linear analysis. They showed that the amplitude of the radial oscillations is damped as the viscosity is increased. These results (see also \citealt{Lubow:2002aa}) were achieved using a 1D approach modelling the disc evolution with a set of linearised equations of motion.

While many hydrodynamical numerical simulation studies of warped discs have been performed \citep[see, for example,][and references therein]{Nelson:1999aa,Raj:2021aa}, only a small number have been focussed on the tilt profile in the wave-like case. Since the steady oscillatory solutions were predicted by \cite{Ivanov:1997aa} and \cite{Demianski:1997aa}, \cite{Nelson:2000aa} presented 3D hydrodynamical simulations including some with $\alpha < H/R$. They did not recover the predicted tilt profiles in their simulations, instead finding the Bardeen-Petterson effect. They argued that nonlinear hydrodynamical effects lead to increased dissipation, precluding the formation of steady radial tilt oscillations. \cite{Lubow:2002aa} explored this problem using a 1D numerical approach. They found steady radial tilt oscillations in their analytical and numerical solutions, and argued that they may be realised in a fluid disc as the wavelength of the tilt oscillations is relatively large ($\lambda \sim R \gg H$). More recently, \cite*{Nealon:2015aa} revisited this problem with 3D hydrodynamical simulations. They present several simulations which exhibit an oscillation of the disc tilt with radius, with a peak between $1-2$ inner disc radii, and a minimum between $2-3$ inner disc radii (see, e.g., their Fig.~4). The properties of the oscillations reported by \cite*{Nealon:2015aa}, including the number of oscillations and their amplitude, are not in agreement with solutions of the 1D linearised equations of motion. This is due to a number of factors including modest resolution and the accompanying non-negligible numerical viscosity that is also present in many previous investigations of low-viscosity warped discs.

In this paper, we present the first 3D hydrodynamical simulations of low-viscosity warped discs around spinning black holes in which the viscosity is isotropic (allowing direct comparison with the available analytical work) and the warp propagation is truly wavelike (with the total viscosity, including numerical viscosity, measured to be such that $\alpha \ll H/R$). We present simulations with a range of parameters to examine several aspects of their evolution. We find that at low initial tilts our simulations exhibit steady radial tilt oscillations that show good agreement with solutions of the 1D linearised equations of motion. Following on from this agreement, we are able to study the impact of non-linear effects at larger initial tilts. We also look at the possibility of disc tearing in the wave-like regime. In many previous works that have explored this effect the magnitude of the numerical viscosity has not been calculated, and we show that once this effect is included it is likely that the simulations reported in the literature to date have been at best in the marginal regime where $\alpha \sim H/R$. We calculate the magnitude of the numerical viscosity in our simulations at the point at which the disc breaks and are able to confirm that the disc is in the wavelike regime with $\alpha < H/R$. In addition, we examine the effect of varying the outer disc radius on the simulation results. 

The layout of the paper is as follows. In Section~\ref{simulations} we present our simulations and results. These are sub-divided into a description of our simulations and their setup (Section~\ref{setup}), simulations and results pertaining to the steady radial oscillatory shape of the disc tilt at small warp amplitudes (Section~\ref{radosc}), results and discussion of the transition from linear to nonlinear propagation of the warp (Section~\ref{nonlinear}), disc tearing in the wavelike regime (Section~\ref{tearing}), and finally the long term evolution of the tilt profile (Section~\ref{longterm}). We provide discussion of our results including their impact on understanding some aspects of observations of accreting black holes in Section~\ref{discussion}, and we present our conclusions in Section~\ref{conclusions}. In the Appendix we provide the numerical methodology for mass injection in SPH simulations of discs that we employ for some of the simulations presented in Section~\ref{longterm}.

\section{Numerical simulations}
\label{simulations}
We present 3D simulations of an initially tilted accretion disc around a Kerr black hole. As we wish to compare with existing analytical work, we require a numerical method in which the viscosity is well-understood and isotropic. Therefore to model the accretion disc, we use the Smoothed Particle Hydrodynamics (SPH) code {\sc phantom} \citep{Price:2018aa}. {\sc phantom} has been used extensively to model accretion discs, including warped discs \citep[e.g.\@][]{Lodato:2010aa,Nixon:2012ac,Nixon:2012ad,Martin:2014aa,Nealon:2015aa,Dogan:2015aa,Raj:2021aa}. The numerical viscosity in SPH is added through explicit terms which are designed to ensure that the Lagrangian is differentiable by keeping the velocity field single-valued; as the numerical viscosity corresponds to explicit terms in the equation of motion it is possible to calculate the magnitude of the effective viscosity that arises from these terms for given disc conditions \citep{Murray:1996aa,Lodato:2010aa,Meru:2012aa}. To model the black hole we make use of post-Newtonian approximations \citep{Nelson:2000aa}, employing the Einstein potential to model the apsidal precession of disc orbits and a gravito-magnetic force term to model the Lense-Thirring (nodal) precession. The methodology is discussed in the {\sc phantom} code paper \citep{Price:2018aa}, and additional detail, including the resulting apsidal and nodal precession frequencies and the orbital shear parameter alongside comparisons with the Kerr solutions, is presented in \cite{Raj:2021aa}.

There are, of course, other numerical methods that are used to simulate misaligned discs around spinning black holes, including GRMHD grid-based codes. The interested reader is referred to \cite{Fragile:2007aa,Henisey:2012aa,Generozov:2014aa,Morales-Teixeira:2014aa,White:2019aa,Liska:2019aa,Liska:2020aa} for more information on these.

\subsection{Simulation Setup}
\label{setup}
For our simulations we set up a disc of $N_{\rm p}$ gas particles, which is initially uniformly inclined by an angle $\beta_0$ to the $z$-axis (with the black hole angular momentum vector parallel to the $z$-axis). The black hole spin is $a = \frac{2}{3}(4-\sqrt{10}) \sim 0.5585$, which was chosen such that the innermost stable circular orbit of the disc is at $R_{\rm in} = 4 R_{\rm g}$ \citep{Lubow:2002aa}. The outer boundary of the disc setup is given by $R_{\rm out}$. We typically use $R_{\rm out} = 40R_{\rm g}$ to ensure a well-resolved disc, but we include simulations with varying $R_{\rm out}$ below.  The initial surface density profile of the disc is given by $\Sigma = \Sigma_0 (R/R_{\rm in})^{-p}(1-\sqrt{R_{\rm in}/R})$ and we use a locally-isothermal equation of state with sound speed $c_{\rm s} = c_{{\rm s,}0} (R/R_{\rm in})^{-q}$. Here, we take $p=1.5$ and $q=0.75$, and the sound speed normalisation is set by the disc scale-height at the inner disc radius, which is $H/R = 0.05$ unless otherwise stated. The disc mass is set to 0.1\% of the black hole mass, although it is worth noting that we do not include self-gravity or the back reaction of the Lense-Thirring precession on the black hole, and thus the disc mass plays no role in the simulation results. Unless stated otherwise, the physical viscosity is set to zero so that we may keep the total simulation viscosity (which includes the numerical viscosity) as low as possible, and ideally in the low-viscosity regime ($\alpha < H/R$). We note that this means that the level of (numerical) viscosity in our simulations is resolution dependent, and thus we expect a modest level of variation in the results with varying resolution. For the numerical viscosity we employ a linear viscosity coefficient, $\alpha_{\rm AV}$, with values between $\alpha_{\rm AV}^{\rm min} = 0.01$ and $\alpha_{\rm AV}^{\rm max} = 1$ determined by a switch \citep{Cullen:2010aa}, and a constant quadratic viscosity coefficient $\beta_{\rm AV} = 2$ \citep[see][for details]{Price:2018aa}. The initial tilt values we use are $\beta_0 = 1^\circ$, $3^\circ$, $10^\circ$, and $30^\circ$. These values are chosen to investigate a range of warp propagation regimes from linear (small warp amplitude) to non-linear (large warp amplitude). Most of the simulations are run for a time of $10^4\,GM/c^3$, which for the parameters given above, corresponds to approximately the time taken for waves travelling at half the sound speed to traverse the disc. This is sufficiently long to ensure the inner disc regions have had time to evolve, but not so long as to be influenced by the (numerical) choice of outer disc boundary.

As discussed above, we typically do not impose a physical viscosity in these simulations as we would like to explore the low-viscosity case. Taking values at $R/R_{\rm in} = 2$ for illustration, the shell-averaged $\alpha_{\rm AV}$ values from the numerical viscosity switch are approximately 0.3, 0.25 and 0.2 respectively, and the shell-averaged smoothing lengths per disc scale-height are 0.52, 0.22 and 0.1. Using the formula derived in the continuum limit by \cite{Meru:2012aa}, we estimate that this results in a Shakura-Sunyaev viscosity in the simulations of $\alpha_{\rm SS}^{\rm AV} = 0.03$, $0.007$ and $0.002$ respectively. Of course, these are values taken at the representative radius of $R/R_{\rm in} = 2$; at larger radii the disc is slightly better resolved so the viscosity is lower there, while at smaller radius the disc surface density decreases strongly and the viscosity is larger there, implying that these values should be taken as an estimate while noting that they are time and spatially dependent.

\subsection{The steady oscillatory disc shape}
\label{radosc}
\begin{figure*}
	\includegraphics[width=0.32\textwidth]{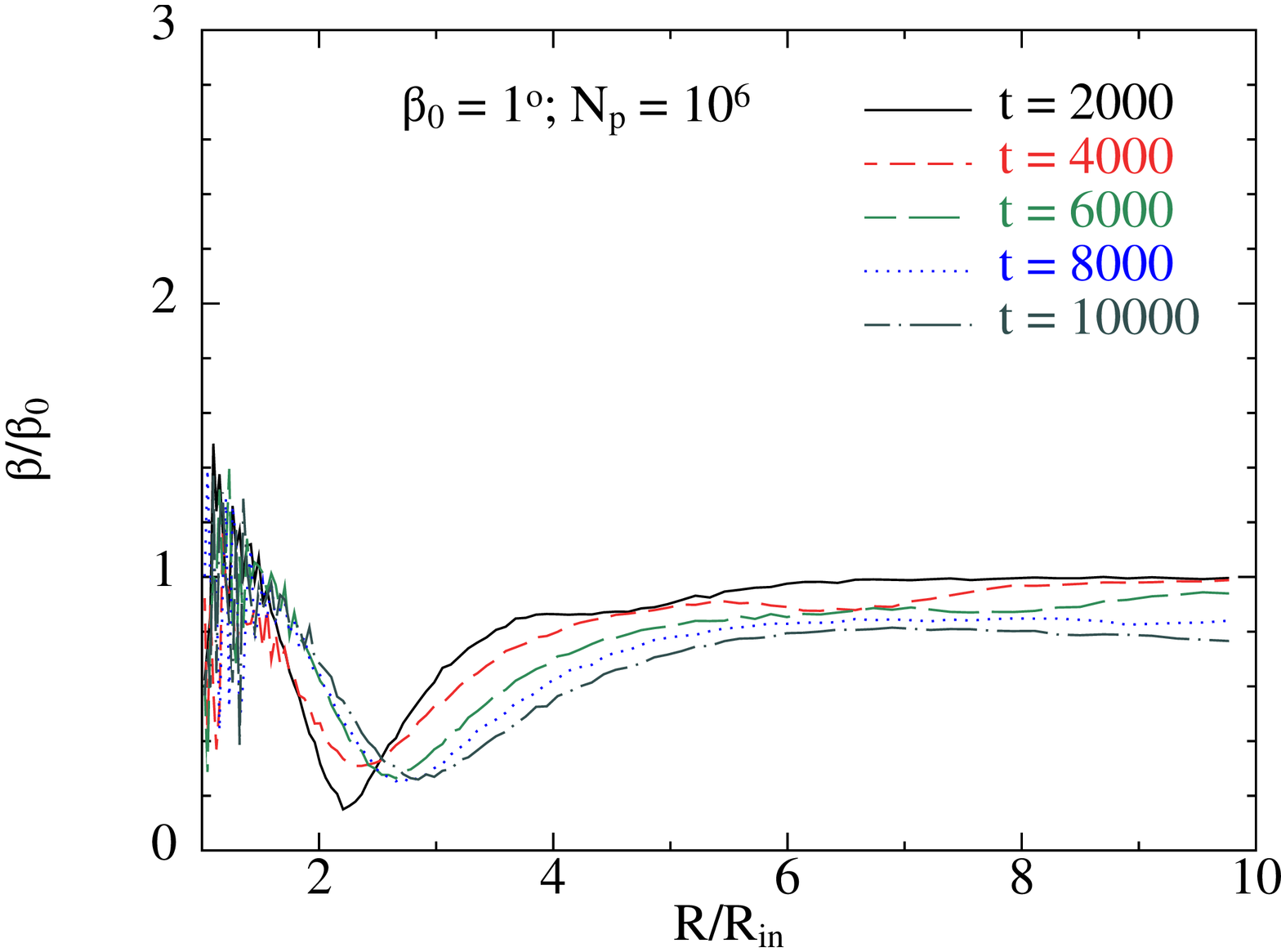}\hfill
	\includegraphics[width=0.32\textwidth]{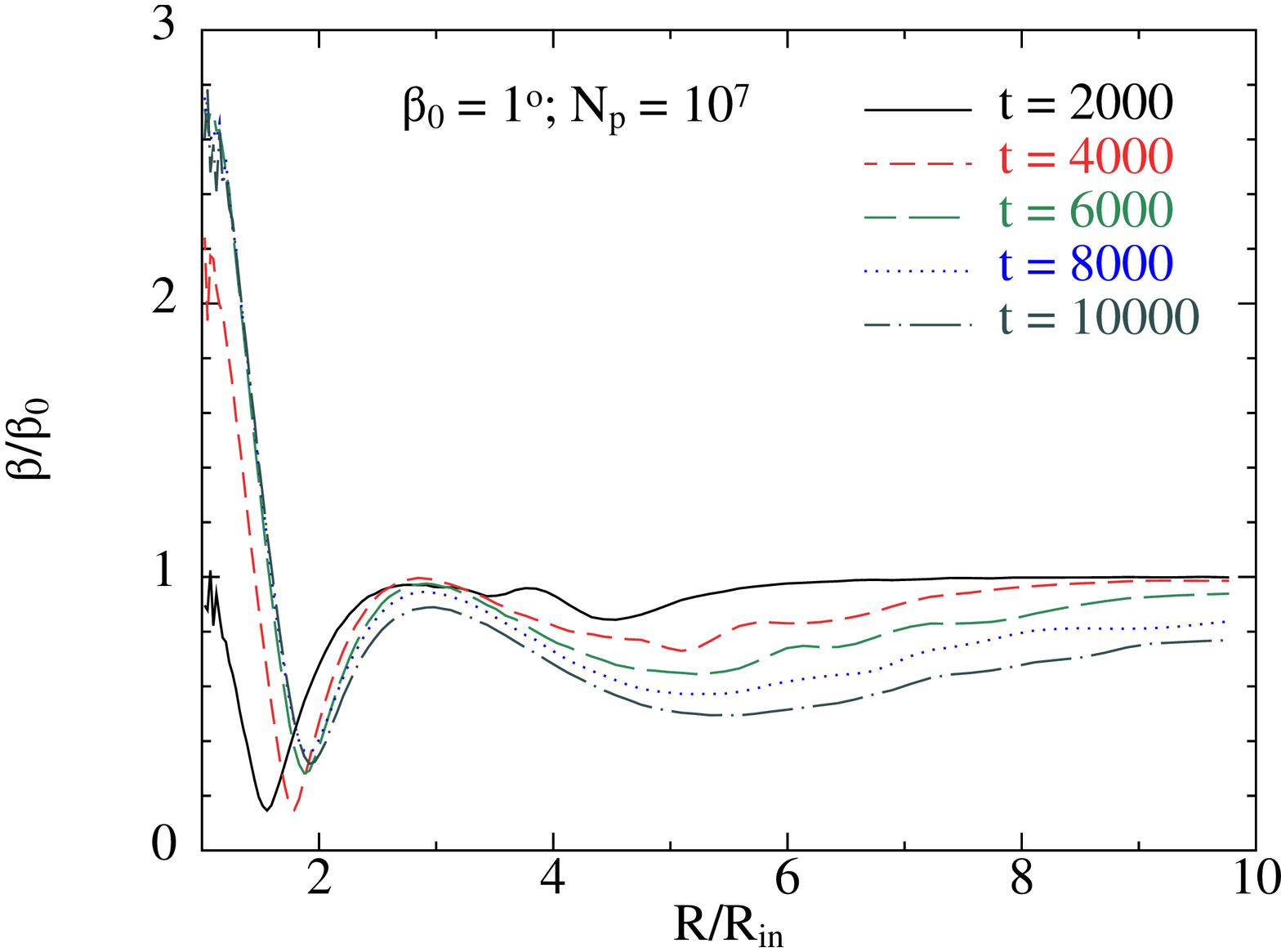}\hfill
	\includegraphics[width=0.32\textwidth]{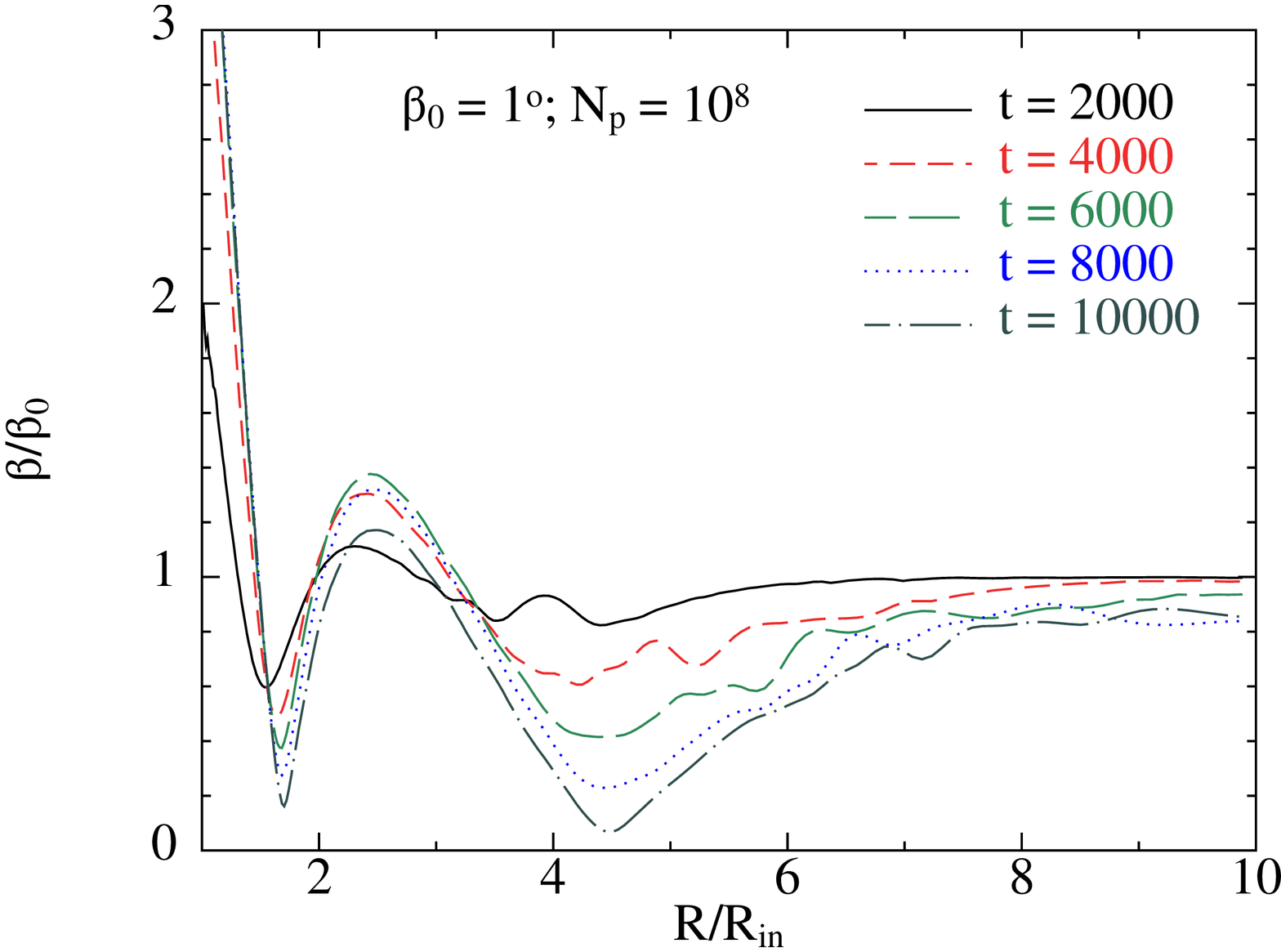}
	\caption{Tilt profiles of the discs that are initially inclined by $\beta_0=1^\circ$. The tilt, $\beta$, is scaled to the initial tilt value of $\beta_0=1^\circ$, and plotted against the radius, $R$, scaled to the inner radius $R_{\rm in}$. Each panel shows the tilt profile at five different times as indicated in the legend, where $t=10^4$ (in units of $GM/c^3$) corresponds to approximately the time taken for a warp wave travelling at half the sound speed to traverse the disc. The left plot shows the disc simulated with $N_{\rm p} = 10^6$, the middle plot shows the disc with $N_{\rm p} = 10^7$, and the right plot shows the disc with $N_{\rm p} = 10^8$. The disc angular semi-thickness is $H/R=0.05$ at $R_{\rm in}$, and this corresponds to $\approx 3^\circ$. Note that as there is no physical viscosity in these simulations, the viscosity is numerical and decreases with increasing resolution. The estimated level of viscosity in each simulation (see text) is $\alpha_{\rm SS}^{\rm AV} = 0.03$, $0.007$ and $0.002$ respectively. }
	\label{Figm1}
\end{figure*}

First we report the results of our default simulation performed with the parameters described above (Section~\ref{setup}). In Fig.~\ref{Figm1} we show the tilt profiles of the simulations with initial inclination of $\beta_0 = 1^\circ$ with the disc modelled with $10^6$ particles (left panel), $10^7$ particles (middle panel) and $10^8$ particles (right panel). In each case the disc quickly establishes a radial oscillatory profile of the disc tilt, and this feature persists for the duration of the simulation. The radial tilt oscillations presented in Fig.~4 of \cite*{Nealon:2015aa} are similar to our lowest resolution simulations (left panel of Fig.~\ref{Fig1}). As the resolution is increased in the middle and right panels, and thus the viscosity correspondingly decreased, the tilt oscillation becomes significantly clearer with a full oscillation of the disc tilt visible for the middle and right panels. We also see that the inner edge of the disc exhibits an increased value of the disc tilt for the high resolution (low viscosity) cases. We attribute this to the conservation of the tilted component of the angular momentum in the propagating warp waves \citep[][see also equation 20 of \citealt{Lubow:2002aa}]{Nixon:2010aa}. As these waves approach the innermost region of the disc, where the surface density approaches zero, the tilt value must increase when the viscosity in negligible. This was not evident in the solutions of \cite{Lubow:2002aa} as they employed a different surface density power-law, and they extended that power-law to $R_{\rm in}$. In our calculations we have the surface density going towards zero at the inner boundary due to the application of a zero-torque boundary condition \citep[see][for a discussion]{Nixon:2021aa}. 

\begin{figure*}
	\includegraphics[width=0.32\textwidth]{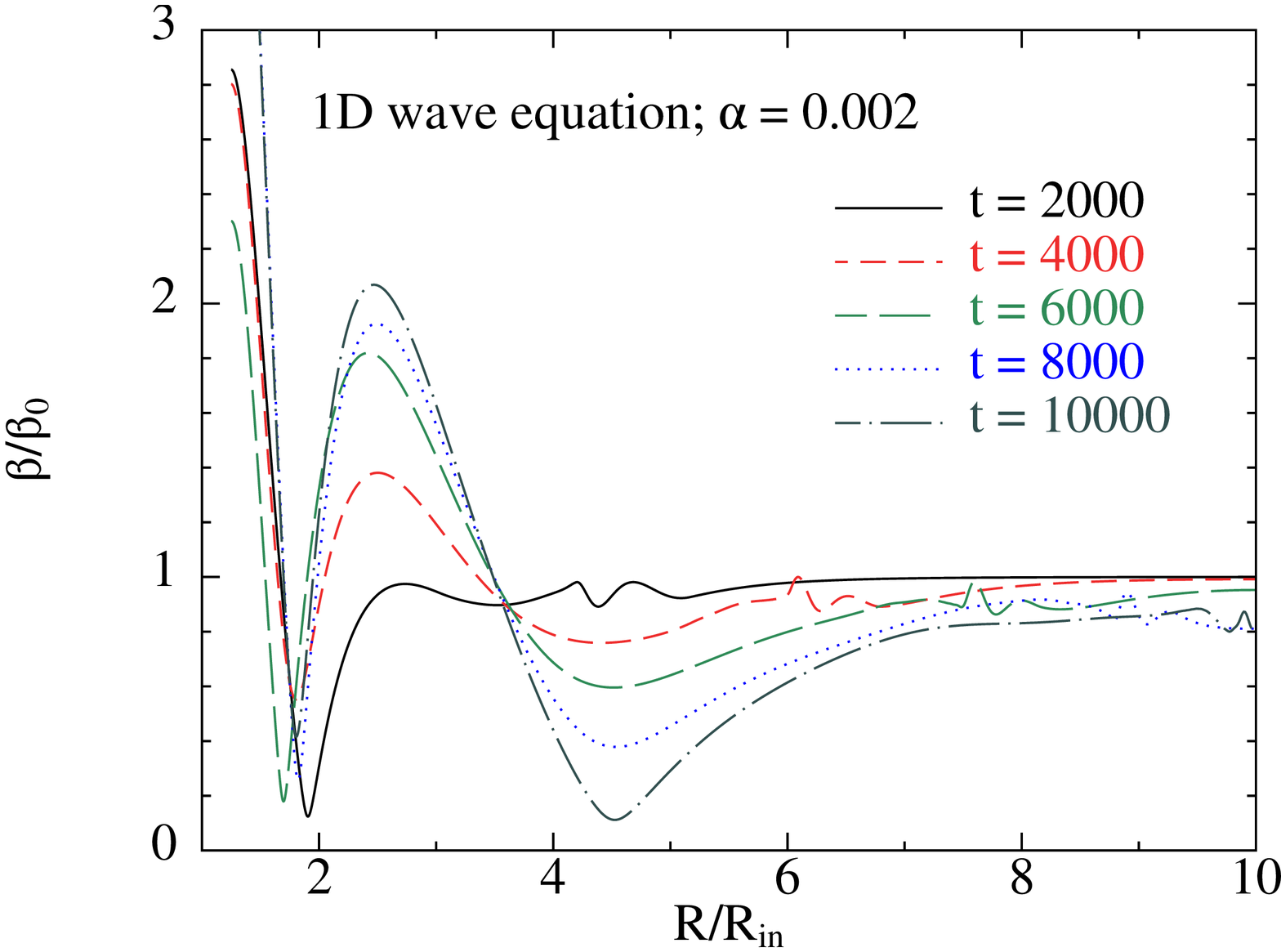}\hfill
	\includegraphics[width=0.32\textwidth]{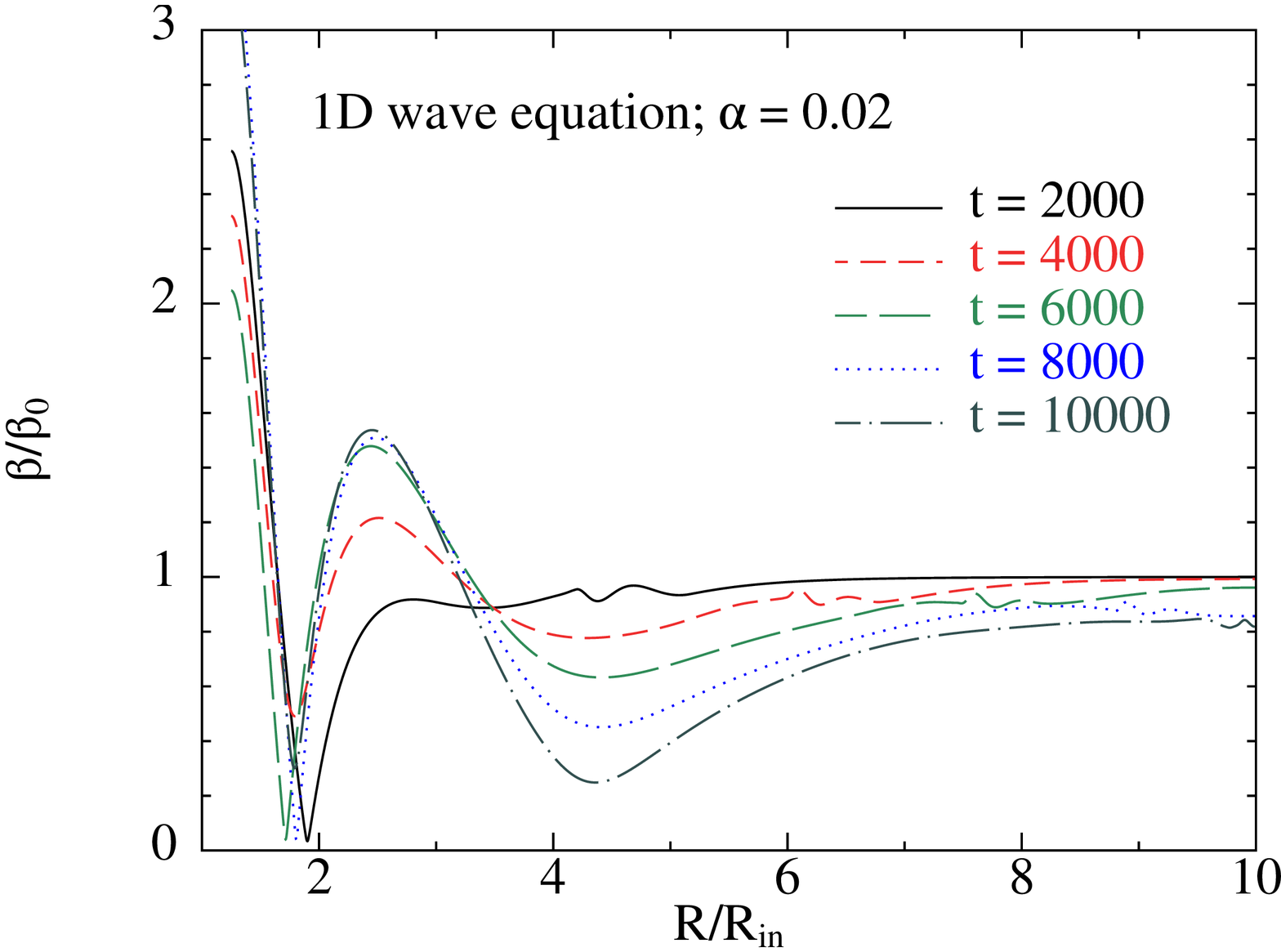}\hfill
	\includegraphics[width=0.32\textwidth]{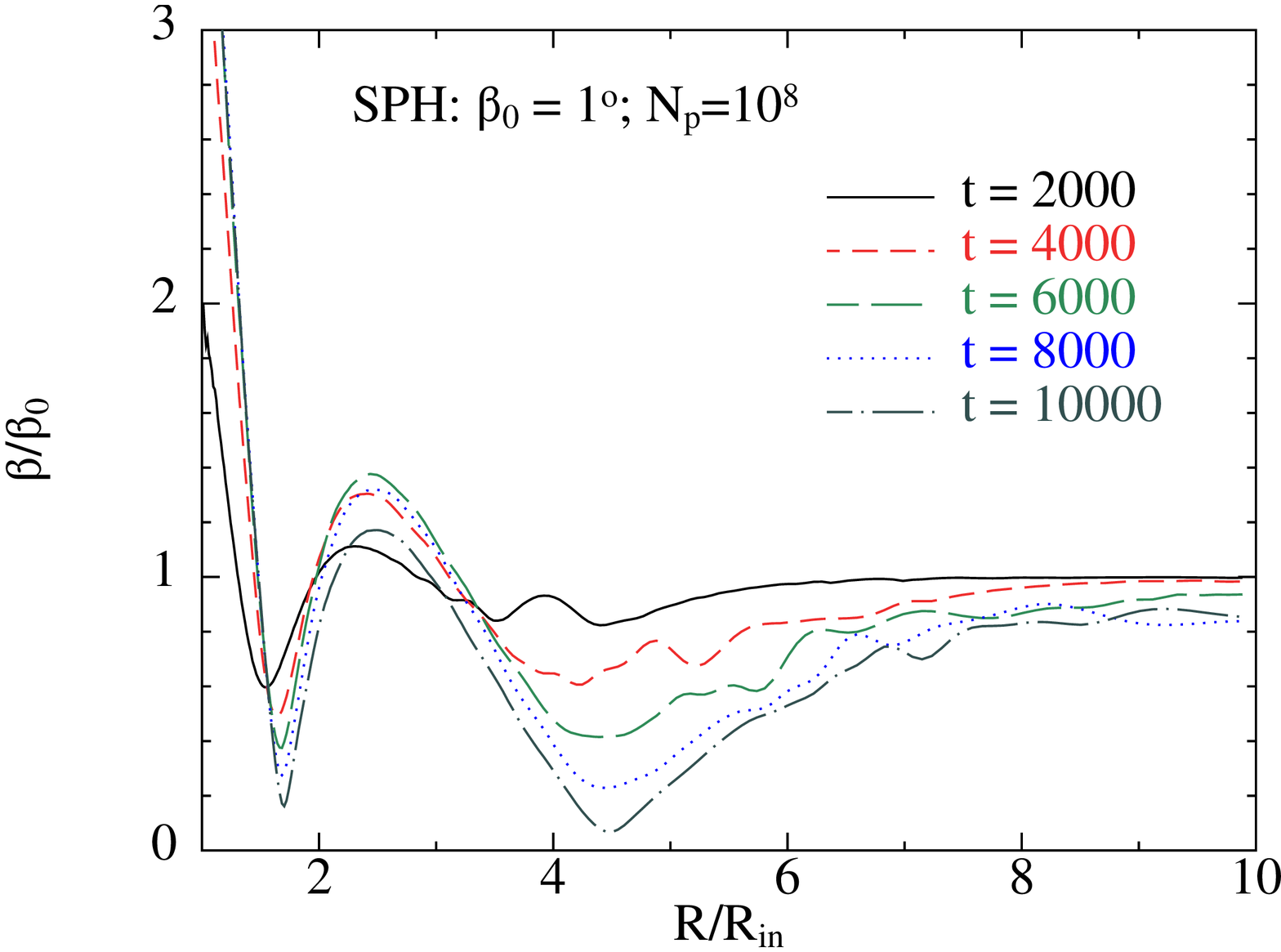}
	\caption{Comparison between solutions to the 1D wave-like equations for a warped disc and 3D hydrodynamical simulation using SPH. The format of the plots is the same as that presented in Fig.~\ref{Figm1} with the disc tilt normalised to the initial disc tilt plotted against the disc radius scaled to the inner disc radius. The different curves correspond to the times given in the legend. The left panel shows the 1D solution with $\alpha=0.002$, the middle panel shows the 1D solution with $\alpha=0.02$, and the right panel shows the SPH simulation with $\beta_0 = 1^\circ$ and $N_{\rm p} = 10^8$ (also shown as the right panel of Fig.~\ref{Figm1}). Differences between the different numerical solutions, and a discussion of the comparison, are given in the text.}
	\label{Fig0}
\end{figure*}

When the tilt angle between the (low-viscosity) disc and the black hole is small, for example $\beta < H/R$ radians ($\approx 3^\circ$ for $H/R = 0.05$), we might expect the disc to behave in the linear regime of wave-like warp propagation \citep{Papaloizou:1983aa,Pringle:1999aa}. That is to say that the disc evolution follows the linearised wave-like equations for a warp \citep{Papaloizou:1995aa,Demianski:1997aa,Lubow:2000aa}. In this case we may predict the disc structure we are expecting from the 3D hydrodynamical simulations by using the numerical method presented in \cite{Lubow:2002aa} for solving the time-dependent 1D wave-like warp equations. In Fig.~\ref{Fig0} we present the tilt profiles from solutions to the 1D equations for two cases, one with $\alpha=0.002$ and one with $\alpha=0.02$, and a comparison with the result of our default 3D hydrodynamical simulation performed with the parameters described above (Section~\ref{setup}) and with $\beta_0 = 1^\circ$ and $N_{\rm p}=10^8$. There are some important differences between the two types of simulations. First, in the 1D calculations we take $\alpha$ to be a constant, while in the SPH simulations the (numerical) viscosity is a function of radius; in this case $\alpha \approx 0.002$ in the body of the disc ($R/R_{\rm in}\gtrsim 2$) and increases due to the surface density profile to $\alpha \approx 0.05$ at $R_{\rm in}$. Secondly, the surface density profile of the 1D calculations is taken to be the input profile (given above) for the SPH simulations, but the surface density profile of the SPH simulation is free to evolve over time. We find that for almost all of the disc the surface density profile in the SPH simulation is unchanged for the duration of the simulation, except for the innermost regions where the surface density turns over to zero at $R_{\rm in}$; here the numerical viscosity is not small and the disc accretes too quickly resulting in the effective disc inner edge moving outwards slightly to $R\approx 1.25R_{\rm in}$. We therefore set the inner boundary of the 1D calculations to be at $1.25R_{\rm in}$. Thirdly, for numerical convenience we take the tilt at the inner edge of the 1D calculations to be zero in the initial conditions; we achieve this by taking the tilt to be zero at the inner boundary and then stepped up to unity with a cosine-bell at a radius of $R/R_{\rm in}=2$ with a bell-width of $R_{\rm in}$ (cf. Section 4.1 of \citealt{Lubow:2002aa}). Finally, the precession frequencies of the disc orbits are modelled in the SPH simulations using a post-Newtonian approximation (described above), while in the 1D calculations we employ those given in equations 14 \& 15 of \cite{Lubow:2002aa}.

In Fig.~\ref{Fig0} we can see that the 1D solutions and the SPH simulation results show similar steady tilt profiles. In particular the location of maxima and minima of the disc tilt are consistent between the two. The disc tilt for $R/R_{\rm in} \gtrsim 4$ is similar between the SPH simulation and the 1D solution with $\alpha = 0.002$ indicating that the numerical viscosity in the SPH simulations is small there. For $R/R_{\rm in} \lesssim 3$ the SPH simulation shows the same features as the 1D solutions, albeit with a reduced amplitude. This reflects the increased numerical viscosity present near the inner edge of the disc (where $\alpha_{\rm SS}^{\rm AV}\approx 0.05$). It is also worth noting that the warp amplitude in the innermost region of the disc $R/R_{\rm in} < 2$ is not small, peaking at a value of $\psi \equiv R\left|\partial\mathbi{l}/\partial R\right| \approx 0.15$,\footnote{Here $\psi$ is the dimensionless warp amplitude which, in general, is a function of radius and time, and $\mathbi{l}$ is the unit tilt vector for the disc which is parallel to the local orbital angular momentum vector.} and thus even for a tilt as small as $\beta_0=1^\circ$ we may expect some nonlinear dissipation in the innermost region of the disc (see Section~\ref{nonlinear} below for more detailed discussion). In both types of solutions we see that the tilt angle becomes large---several times the initial tilt value---near $R_{\rm in}$. We have confirmed (not depicted) that the increase in disc tilt at the inner edge is diminished if the surface density follows a power-law to the inner radius as assumed by \cite{Lubow:2002aa}. It is worth noting that even when the surface density power-law is extended to $R_{\rm in}$ it is possible for the inner disc tilt to exceed the outer disc tilt depending on the surface density and sound speed power-laws (cf. equation 20 of \citealt{Lubow:2002aa}).

From these simulations we can see that, in low-viscosity discs, steady radial oscillations of the disc tilt can be formed in fluid discs, and further that the results of hydrodynamic simulations are in close agreement with solutions to the time-dependent wave-like warped disc equations when the tilt (and hence warp amplitude) are small. The fact that the inner disc can be significantly more tilted than the outer disc may have important implications for the observational properties of these discs, and we return to this in the discussion. Finally, we also note from these figures that while the tilt oscillations are present and have wavelength $\lambda \sim R$, it is also possible to detect propagating waves with $\lambda \sim H$ in the tilt profiles of the SPH simulations (present only at high resolution for which the viscosity is significantly smaller than the disc angular semi-thickness). These features appear to be continuously driven from the inner disc region, where the warp amplitude is largest. We observe that the inner disc tilt (at $R/R_{\rm in}\lesssim 1.5$) undergoes small amplitude oscillations in time throughout the simulation due to the low numerical viscosity (and lack of a physical viscosity to damp such motions). The outward-propagating short-wavelength tilt waves may be driven by this time-dependent oscillation of the inner disc tilt. We shall investigate these effects in a later paper.

\subsection{On the transition from linear to nonlinear propagation}
\label{nonlinear}
\begin{figure*}
	\includegraphics[width=0.32\textwidth]{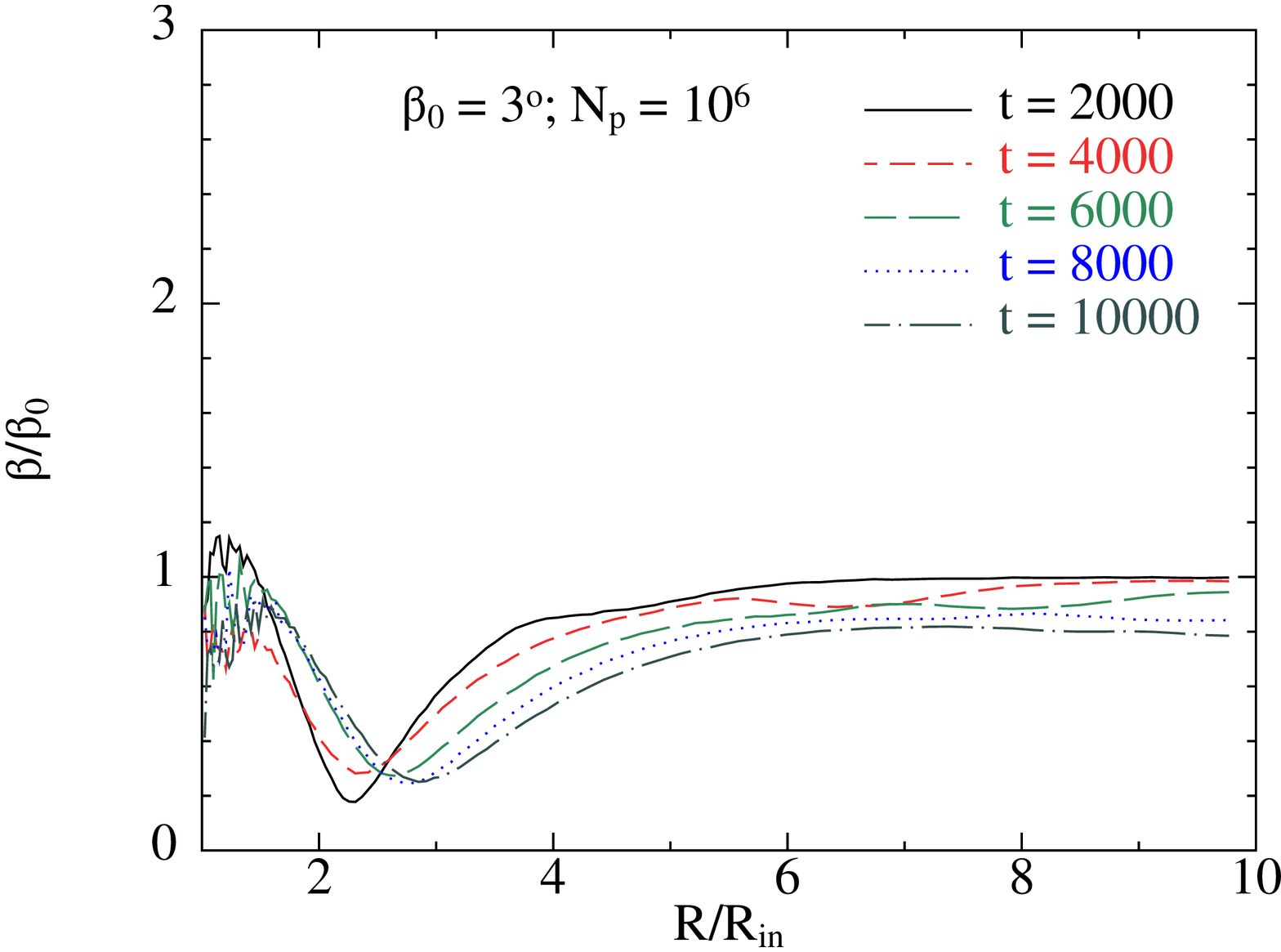}\hfill
	\includegraphics[width=0.32\textwidth]{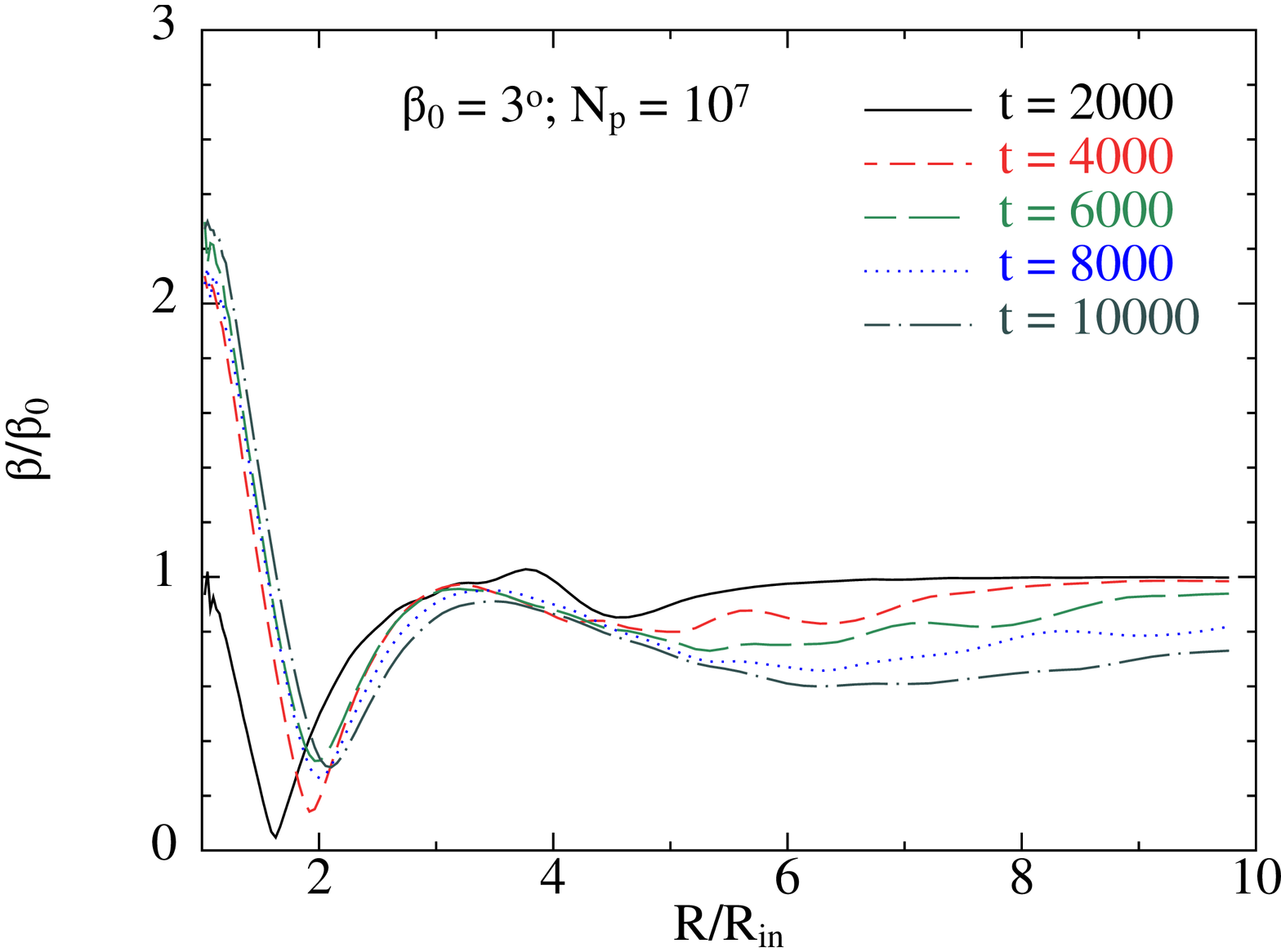}\hfill
	\includegraphics[width=0.32\textwidth]{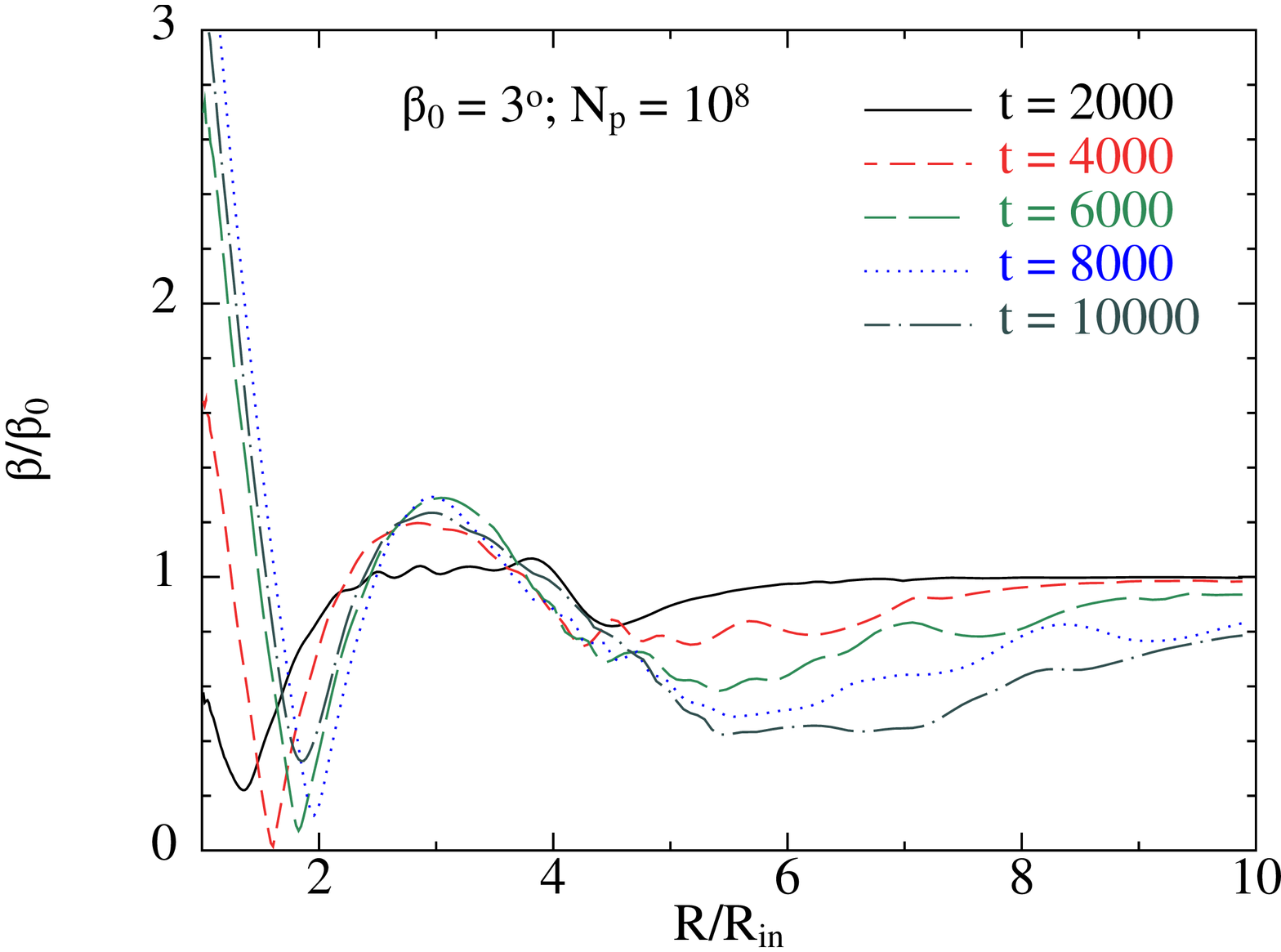}
	\caption{Tilt profiles of the discs that are initially inclined by $\beta_0 = 3^\circ$, for comparison with the $\beta_0 = 1^\circ$ simulation presented in Fig.~\ref{Figm1}. The format of the plots is the same as presented in Fig.~\ref{Figm1}. The disc angular semi-thickness is $H/R=0.05$ at $R_{\rm in}$, and this corresponds to $\approx 3^\circ$. For $\beta_0 = 3^\circ$ we see very similar tilt profiles as shown in the $\beta_0=1^\circ$ case when $N_{\rm p} = 10^6$ and $10^7$. However, when $N_{\rm p}=10^8$ the $\beta_0=3^\circ$ simulation exhibits a solution that indicates the presence of additional local dissipation compared to the corresponding $\beta_0=1^\circ$ simulation. It is worth noting that the inclinations in these two cases are small enough that the disc surface density, and thus the numerical viscosity, is not strongly affected by the presence of the warp. The main difference between the two cases is the amplitude of the warp, which is larger for larger $\beta_0$. Thus we can conclude that $\beta_0 = 3^\circ$ is large enough for the disc evolution to be affected by nonlinear dissipation.}
	\label{Fig1}
\end{figure*}

\begin{figure*}
	\includegraphics[width=0.32\textwidth]{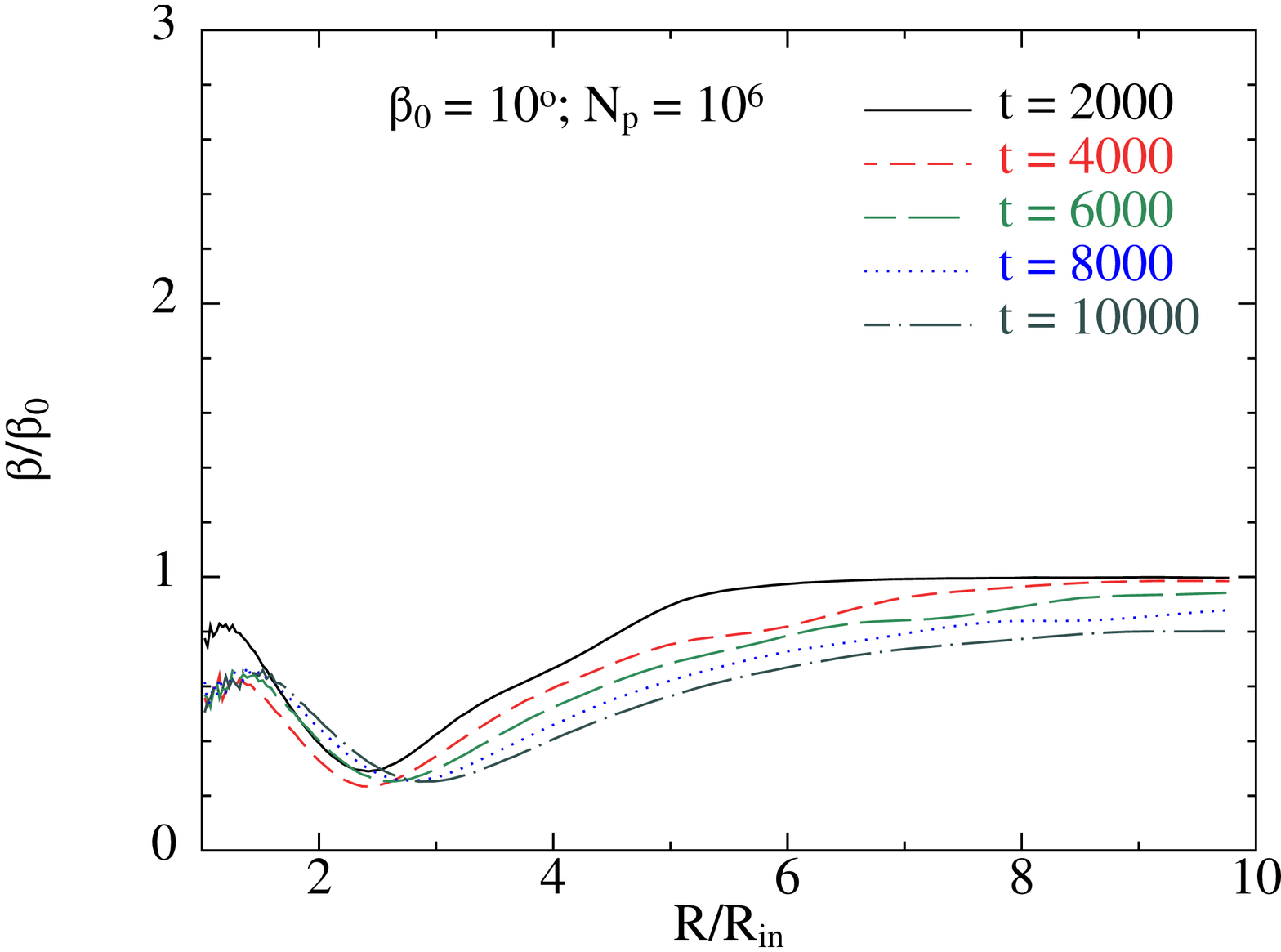}\hfill
	\includegraphics[width=0.32\textwidth]{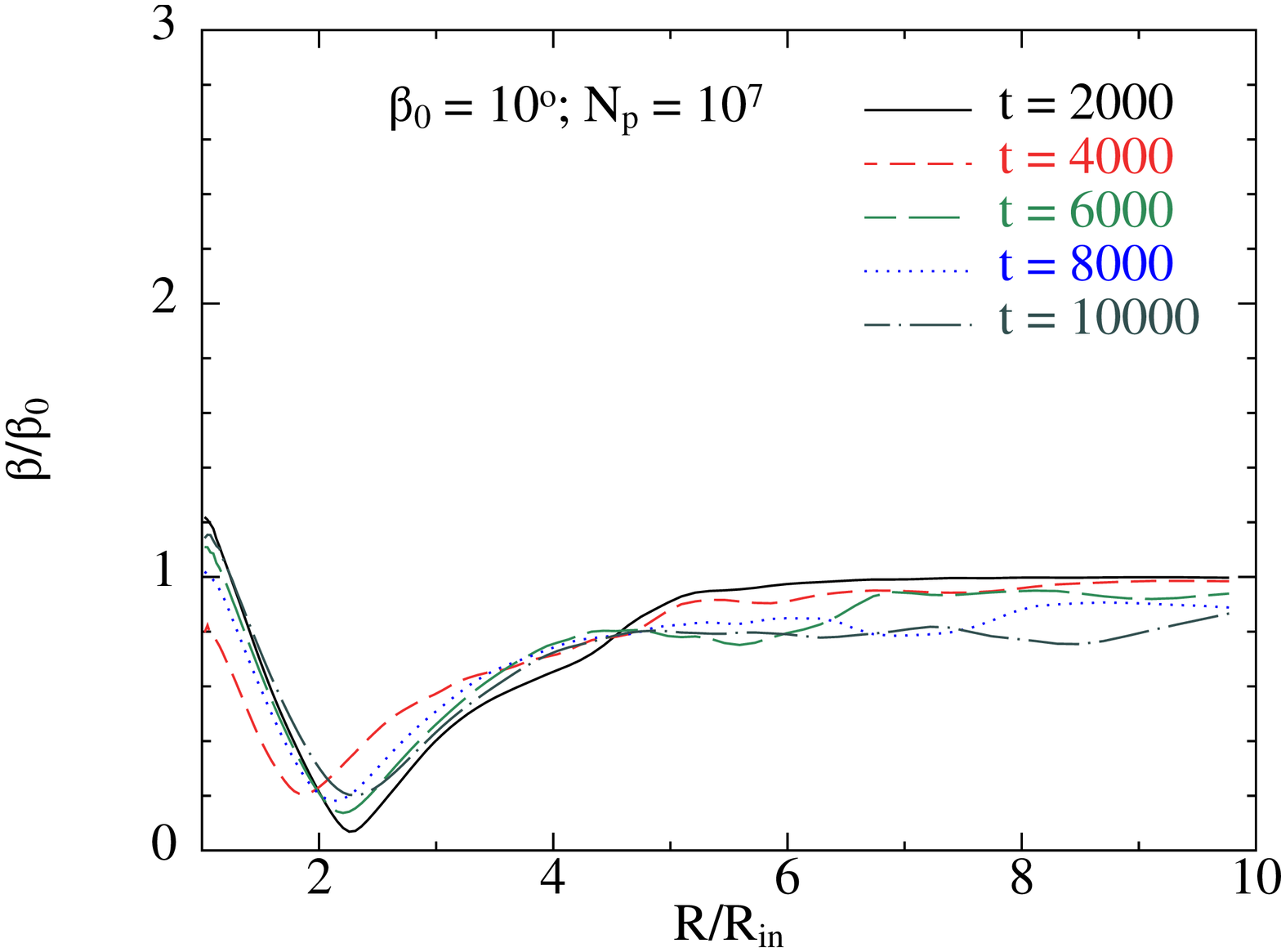}\hfill
	\includegraphics[width=0.32\textwidth]{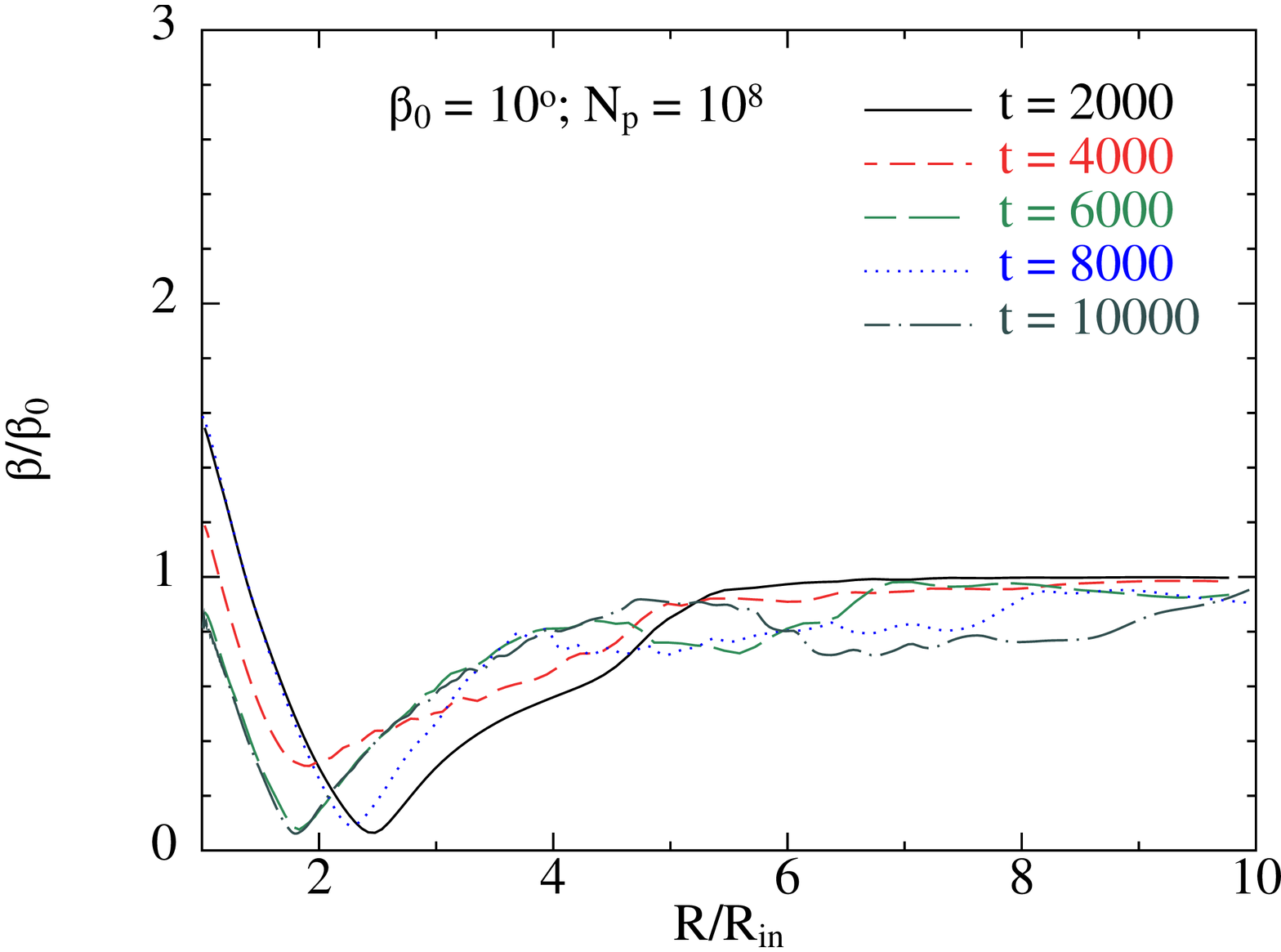}
	\caption{Tilt profiles of the discs that are initially inclined by $\beta_0 =10^\circ$, for comparison with the $1^\circ$ tilt simulation presented in Fig.~\ref{Figm1} and the $3^\circ$ tilt simulation presented in Fig.~\ref{Fig1}. The format of the plots is the same as presented in Fig.~\ref{Figm1}. The disc angular semi-thickness is $H/R=0.05$ at $R_{\rm in}$, and this corresponds to $\approx 3^\circ$. At $\beta_0 = 10^\circ$ we see significant deviation from the solutions at lower inclinations. It is worth noting that, again, the inclinations in these two cases are small enough that the disc surface density, and thus the numerical viscosity, is not strongly affected by the presence of the warp. In this case variations in the surface density, and thus the local resolution, lead to changes in the numerical viscosity by less than a factor of two. The main difference between the two cases is the amplitude of the warp, which is larger for larger $\beta_0$. Thus we can conclude that $\beta_0 = 10^\circ$ is large enough for the disc evolution to be strongly affected by nonlinear effects as the disc shape is no longer consistent with the shape presented for $\beta_0 = 1^\circ$ or $3^\circ$.}
	\label{Fig3}
\end{figure*}

For comparison with the $\beta_0 = 1^\circ$ simulations presented in Fig.~\ref{Figm1}, we plot in Figs~\ref{Fig1} \& \ref{Fig3} the corresponding data for the $\beta_0 = 3^\circ$ and $10^\circ$ simulations respectively. The warp amplitudes, $\psi \equiv R\left|\partial\mathbi{l}/\partial R\right|$, vary across the disc, but we typically see that the warp amplitude is larger for larger values of $\beta_0$ as expected. In the inner regions ($R/R_{\rm in}\lesssim 3$), a quasi-steady tilt profile has formed. At $R/R_{\rm in}=2$ the warp amplitude (measured at a time of $t=8000\,GM/c^3$) is 0.065, 0.25 and 0.33 for the $\beta_0 = 1^\circ$, $3^\circ$ and $10^\circ$ simulations respectively. At smaller radii the warp amplitudes are larger by a factor of $2-3$, and at larger radii the warp amplitudes are smaller (except where there are short-wavelength warp waves propagating through the disc; which are visible in the tilt profiles). It is generally expected that once the warp amplitude becomes large then nonlinear effects become increasingly important \citep[see, for example,][]{Pringle:1999aa,Ogilvie:2006aa}. For wavelike propagation of the disc warp, nonlinear effects may arise from dissipation due to the extraction of energy from the local shear flow in the warp. This may be associated with shear instabilities such as Kelvin-Helmholtz \citep{Kumar:1993aa}, and shocks resulting from supersonic shearing motions when the warp amplitude is sufficiently large \citep{Pringle:1999aa}. Additionally wavelike warp propagation may be subject to nonlinear effects that are not directly dissipative in nature, for example the parametric instability which may occur when the viscosity is sufficiently small \citep{Gammie:2000aa,Ogilvie:2013ab} and nonlinear dispersion of the waves may also occur at sufficient amplitude \citep[Section 7.1 of][]{Ogilvie:1999aa}. We can therefore expect that for small tilts ($\beta_0 < H/R$) and modest resolution (implying modest numerical viscosity; $0 < \alpha < H/R$) we should recover linear behaviour; that is disc evolution that is independent of the initial tilt value. Evidence for this may be found by comparing the tilt profiles for the disc with $\beta_0 = 1^\circ$, $N_{\rm p}=10^7$ (middle panel of Fig.~\ref{Figm1}) with $\beta_0 = 3^\circ$, $N_{\rm p}=10^7$ (middle panel of Fig.~\ref{Fig1}). However, we can see that comparing the same simulations but at $N_{\rm p}=10^8$, the evolution is no longer the same; the evolution for $\beta_0=1^\circ$ recovers a stronger radial oscillation of the disc tilt that agrees more closely with the solutions of the linearised warp wave equations (Fig.~\ref{Fig0}). Therefore we suspect that at $\beta_0=3^\circ$ the highest resolution simulation has a sufficiently small viscosity (and sufficiently large warp amplitude) that there is additional dissipation of the warp due to nonlinear effects, while in the $\beta_0=1^\circ$ case the warp amplitude is too small for nonlinear dissipation to present in a noticeable manner.

For the $10^\circ$ simulation (Fig.~\ref{Fig3}) it is clear that the behaviour is not linear as the results are quite different, particularly at high resolution, from the simulations with lower initial tilts. For this case, at all resolutions, the results are similar to the low-resolution results at lower tilts; i.e.\@ the right panel of Fig.~\ref{Fig3} is similar to the left panel of Fig.~\ref{Fig1} and the left panel of Fig.~\ref{Figm1}. This indicates that at $10^\circ$ the disc is subject to significant additional dissipation in regions of large warp amplitude. It is worth noting that we can rule out with some confidence the possibility that variations in the numerical viscosity are responsible for these changes as there is almost no change in surface density profile (and thus $\left<h\right>/H$) between the $\beta_0=1^\circ$ and $\beta_0=3^\circ$ simulations, and only a modest change seen in the $\beta_0=10^\circ$ simulation leading to at most a factor of $\lesssim 2$ increase in the effective viscosity from the numerical viscosity at any radius; this is insufficient to drive the observed differences (cf. the small difference between the left and middle panels of Fig.~\ref{Fig0}). 

For $\beta_0=30^\circ$ we find that the disc behaviour is dominated by different nonlinear dynamics, namely disc tearing. In this case, the disc is unstable to breaking into discrete planes which can precess effectively independently \citep{Nixon:2012ad,Dogan:2018aa}. We discuss this case in more detail in the next section.

\subsection{Disc tearing in the wave-like regime}
\label{tearing}
\subsubsection{Results in previous work}
\cite{Larwood:1996aa} and \cite{Larwood:1997aa} were the first to present numerical simulations that exhibit a break  between different regions of a warped disc. In these works the disc warp is induced by the gravitational field of a binary companion. In \cite{Larwood:1996aa} the authors callibrate the numerical viscosity in their simulations by comparing the evolution of the surface density of a simulated planar disc to the expected evolution from integrating the one dimensional diffusion equation for a disc. They find that the viscosity is given approximately by $\alpha \sim 0.03$ at the outer disc edge. The simulation in their paper (model 9) which exhibits a disc break has $H/R = 1/30$ (constant through the disc), and thus $\alpha \gtrsim H/R$ in the body of the disc (their disc models are such that $\nu$ is approximatly constant with radius, such that $\alpha \propto r^{-1/2}$; cf. eqn~9 in \citealt{Larwood:1997aa}). The authors also report (not depicted) severe disruption of the disc in model 10, with in this case $H/R = 1/50$, and thus firmly in the diffusive regime. In \cite{Larwood:1997aa} a disc break is shown for model 13, in which the disc is thicker with $H/R = 1/20$. However, their equation~9 gives that the disc has $\alpha \gtrsim H/R$ for the body of the disc, with $\alpha \approx H/R$ only near the outer disc edge. So in both of these papers there is no clear evidence for instability of the disc in the wave-like regime where $\alpha < H/R$.

Disc tearing, where a disc warp is driven by forced precession that causes an unstable region to break apart, was explored in detail by \citet[][for discs around black holes]{Nixon:2012ad} and \citet[][for circumbinary discs]{Nixon:2013ab}. However, these works focussed on the diffusive regime with $\alpha > H/R$. \cite{Facchini:2013aa} present an investigation into the propagation of warps in wave-like, circumbinary discs with a focus on comparing the results of SPH simulations to the results of integrating the linearised equations of motion derived by \cite{Lubow:2000aa}. In one simulation \citet[][Section 6.3.2]{Facchini:2013aa} report a disc break for a simulation with $\alpha = 0.05$ and $H/R = 0.1$, where the quoted value of $H/R$ applies to the inner edge of the disc. They also note that they have performed the same simulation with $\alpha = 0.01$ and that the results are equivalent. The break shown in their simulation occurs approximately at a radius of $x=5$ (their Fig.~16) at which, for their choice of sound-speed profile (yielding $H/R \propto (R/R_{\rm in})^{-0.25}$), the corresponding disc angular semi-thickness is $H/R \approx 0.06-0.07$. This means that the physical viscosity, which they employ via the direct Navier-Stokes viscosity formalism \citep[described in][]{Lodato:2010aa}\footnote{In 2018 (8th August, commit ID: d9e286a) one of the present authors, after testing the physical viscosity terms in {\sc phantom}, included a reduction in the time-stepping associated with the physical viscosity terms by a factor of 0.4. This was required to ensure convergence of the simulations with increasing resolution. Prior to this change lower resolution could, in some instances, yield a larger than desired physical viscosity. The same author speculates that this time-stepping issue was the cause of the small discrepancy between input and fitted viscosity parameter, particularly at low viscosity, of the direct Navier-Stokes viscosity method in the numerical tests presented in \citet[][Fig. 4]{Lodato:2010aa}.}, is not much smaller than $H/R$. \cite{Facchini:2013aa} also include a numerical viscosity with $\alpha_{\rm AV}\approx 0.5$\footnote{Specifically they used the \cite{Morris:1997aa} switch with $\alpha_{\rm AV}^{\rm min} = 0.01$ and $\alpha_{\rm AV}^{\rm max} = 0.5$. At the resolution employed, and for such large amplitude warps, the authors would expect the {\sc phantom} code at that time to have most of the particles spend most of their time with $\alpha_{\rm AV}\approx\alpha_{\rm AV}^{\rm max}$.} and $\beta_{\rm AV}=2$. To calculate the physical viscosity implied by these numbers we need to know the shell-averaged smoothing length per disc scale height, $\left<h\right>/H$, and then use, for example, equations 6 \& 7 in \cite{Meru:2012aa}. \cite{Facchini:2013aa} do not report the resolution used in their simulations in terms of $\left<h\right>/H$, instead only providing the number of particles. However, it seems reasonable to expect that the numerical viscosity was not small compared to the physical viscosity employed---the resolution at which the numerical viscosity would yield a physical viscosity of $\alpha=0.05$ for the simulation parameters given above is\footnote{This value can be compared to the values indicated in Fig.~1 of \cite{Facchini:2018aa}.} $\left<h\right>/H \approx 0.6$---and therefore it is unlikely that their simulation was in the wave-like regime, particularly when the depression of the surface density that accompanies the strong warping prior to the disc breaking is taken into account.

More recently \cite*{Nealon:2015aa} explored the Bardeen-Petterson effect in SPH simulations of tilted discs around a spinning black hole. They present simulations with $H/R=0.05$ at the inner edge, and employ a Shakura-Sunyaev viscosity (modelled by the scaled artificial viscosity method; Section 3.2.3 of \citealt{Lodato:2010aa}) with $\alpha = 0.01$ and $0.03$. They report disc tearing at a radius of $\approx 10R_{\rm in}$, at which $H/R \approx 0.028$. Thus we might conclude that the higher viscosity simulations have $H/R \sim \alpha$ and the lower viscosity simulations have $\alpha < H/R$. However, we need to account for two effects. First in their Fig.~9, they show that $\left<h\right>/H$ has increased by a factor of $2-3$ in the region where the break occurs; in their methodology the effective Shakura-Sunyaev viscosity parameter is linearly proportional to $\left<h\right>/H$. Second, there is an additional contribution to the total simulated viscosity from the $\beta_{\rm AV}$ term. Taking these effects into account we find that the level of viscosity in the simulations presented by \cite*{Nealon:2015aa} corresponds to $\alpha \gtrsim H/R$.

Perhaps the most compelling evidence so far for instability in the wave-like regime can be found in \cite{Fragner:2010aa}. They present a simulation (labelled 6a) which has $H/R = 0.01$ and $\alpha = 0.005$, and they find that the outer part (of their circumprimary disc) breaks from the rest of the disc due to precession induced by a companion. However, while the authors discuss numerical viscosity there is no quantitative measure given. The authors note that their grid-based hydrodynamic code has a relatively low numerical viscosity compared to those SPH schemes that had been used previously, and they show (see their Fig.~1 \& 2) impressive agreement between their numerical results at small inclination angles and the evolution determined by the linearised wave equations for a warp (see Fig.~2 of \citealt{Nealon:2015aa} for similar results achieved with the {\sc phantom} SPH code). However, they also note that for higher inclination angles their code behaves more diffusively (see also the Appendix in \citealt{Sorathia:2013aa}). This may result for physical reasons, such as supersonic shearing motions at larger warp amplitudes as noted by \citet[][see also \citealt{Nelson:1999aa}]{Fragner:2010aa}, but may also indicate an enhanced numerical viscosity in grid-based codes for larger warp amplitudes \citep[and that the numerical viscosity in this case is significantly anisotropic;][]{Sorathia:2013aa}. For the simulations presented by \cite{Fragner:2010aa} a numerical viscosity that yields an effective $\alpha \sim 0.005$ would be enough to imply that $\alpha \approx H/R$ in their simulation that exhibited a broken disc.

More recently, other papers have presented simulations with reported physical viscosity parameters that are smaller than the reported disc angular semi-thickness (see, for example, \citealt{Facchini:2018aa}; \citealt{Nealon:2020ab}), but in general similar arguments to those provided above apply and once numerical viscosity has been accounted for the simulations typically have $\alpha \gtrsim H/R$.\footnote{One possibility to significantly increase the local resolution without significantly increasing the computational cost is to perform local (``shearing box'') simulations and these have been employed to explore detailed internal dynamics in warped disc \citep[see, for example,][]{Ogilvie:2013ab,Deng:2021aa}, but such simulations cannot be directly applied to determine the global behaviour of the disc with regards to disc breaking or radial tilt oscillations.} Given this, we conclude that in previous work there has not been any strong evidence presented for the breaking instability of the warp to occur in the wave-like case. 

\subsubsection{Results presented here}
\begin{figure}
	\centering\includegraphics[width=\columnwidth]{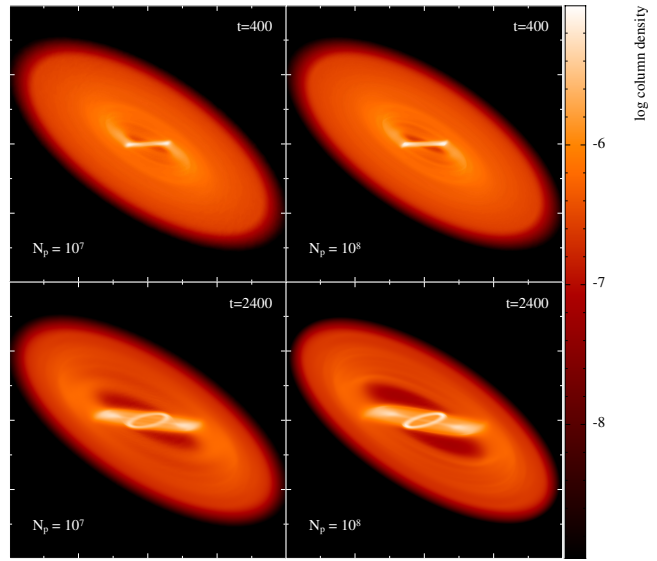}
	\caption{Column density plots of the discs that are initially inclined by $30^\circ$. The left panels correspond to the simulation with $N_{\rm p} = 10^7$, and the right panels correspond to $N_{\rm p}=10^8$. The top panels correspond to a time of $400\,GM/c^3$ at which the disc first exhibits a clear break, and the bottom panels correspond to a time of $2400\,GM/c^3$ which is the latest time at which we have data for the $N_{\rm p}=10^8$ simulation at this inclination. There are only minor differences between the two resolutions (see also Fig.~\ref{FigA} below). The innermost ring precesses throughout the simulation and the tilt of the innermost regions is variable with time exhibiting oscillations and some alignment; by the end of the $N_{\rm p}=10^7$ simulation the innermost ring is misaligned by approximately $5-10^\circ$. As time proceeds a second ring, that is more radially extended than the first, is broken off and this is visible in the lower panels.} 
	\label{FigZ}
\end{figure}

We present in Fig.~\ref{FigZ} the disc structures for the simulations with $R_{\rm out}=40R_{\rm g}$, $\beta_0 = 30^\circ$ and $H/R=0.05$ modelled with $N_{\rm p} = 10^7$ (left panels) and $N_{\rm p}=10^8$ (right panels) at times of $400\,GM/c^3$ (top panels) and $2400\,GM/c^3$ (bottom panels). We chose these times as at $400\,GM/c^3$ the first clear break has developed in the disc at $R\approx 2R_{\rm in} (= 8R_{\rm g})$, and $2400\,GM/c^3$ is as far as we were able to run the $N_{\rm p}=10^8$ simulation at this inclination with available computing resources. It is clear that by comparing the left  and right panels that the increase in resolution has made little difference to the disc evolution (see also Fig.~\ref{FigA} below) indicating that the disc tearing behaviour presented here is numerically converged.\footnote{It is worth remarking that \cite{Raj:2021aa} found, contrary to the prediction of a local stability analysis \citep{Dogan:2018aa}, that the growth rates of the instability in unstable disc regions were relatively insensitive to the level of the viscosity. Our results support this finding, but additional investigation is required to determine if this is always the case.} We also see that the disc breaks into multiple rings from the inside outwards \citep{Nixon:2012ad,Raj:2021aa}.

To explore the disc structure at the onset of disc tearing in detail we plot in Fig.~\ref{FigA} the disc properties at a time of $400\,GM/c^3$. We plot the surface density $\Sigma$ (top left), the resolution $\left<h\right>/H$ (top right), the tilt angle scaled by the initial tilt $\beta/\beta_0$ (middle left), the twist angle $\gamma$ (middle right), the warp amplitude $\psi$ (bottom left), and the effective Shakura-Sunyaev viscosity parameter arising from the numerical viscosity $\alpha_{\rm SS}^{\rm AV}$ (bottom right). In each case we present the results for $N_{\rm p}=10^7$ as a black solid line, and the results for $N_{\rm p} = 10^8$ as a red dashed line. It is clear that the disc structure is essentially the same at both resolutions, indicated by the similarity of $\Sigma$, $\beta$, $\gamma$, and $\psi$. We can also see the expected increase in resolution indicated by the corresponding drop in the value of $\left<h\right>/H$; increasing the particle number by a factor of 10 leads to a decrease in $\left<h\right>$ by a factor of $\approx 2.15$. This increase in resolution leads to a corresponding drop in the viscosity arising from numerical viscosity. We mark on the bottom right panel of Fig.~\ref{FigA}, with a dotted blue line, the line corresponding to $\alpha_{\rm SS}^{\rm AV} = H/R (=0.05)$. From this we can see that, at the point at which the disc first exhibits a clear break (evident from the sharp spike in $\psi$ at $R\approx 2R_{\rm in}$), the disc with $N_{\rm p}=10^7$ has snuck into the diffusive regime with $\alpha \gtrsim H/R$ at $R\approx 2R_{\rm in}$. However, we can also see that this is not the case for the $N_{\rm p}=10^8$ simulation where instead $\alpha < H/R$ in this region when the disc breaks. From the results presented in Figs.~\ref{FigZ} \& \ref{FigA} we can conclude that the disc tearing behaviour is numerical converged and that physically this behaviour can manifest in wavelike warped discs when $\alpha$ is smaller than, but of the order of, the disc angular semi-thickness.

We speculate here that the disc tearing seen in the wave-like warp propagation regime is inherently the same instability seen in the diffusive warp propagation regime \citep[explored by][]{Dogan:2018aa}. We speculate that this comes about because as the warp amplitude is increased the local dissipation rate in the warp increases, and hence the effective viscosity there may become larger than the disc angular semi-thickness (cf. Section~\ref{nonlinear}). Confirming these speculations requires a more detailed investigation than we are able to present here, and along the lines of that presented in \cite{Raj:2021aa} but for discs that when $\psi \approx 0$ are in the wave-like regime with $\alpha \ll H/R$.

\begin{figure}
	\centering\includegraphics[width=\columnwidth]{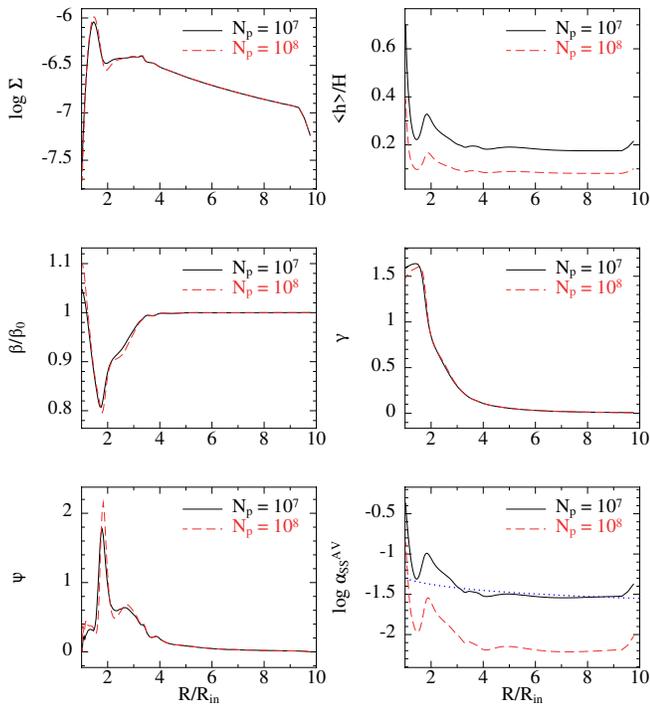}
	\caption{The disc structure for the $\beta=30^\circ$ simulations corresponding to the top panels of Fig.~\ref{FigZ} which show the time at which the disc first exhibits a clear break. The break occurs at $R\approx 2R_{\rm in}$, and is most clearly identified by the sharp spike in warp amplitude (bottom left panel). In each panel the black solid line shows the data for the $N_{\rm p}=10^7$ simulation, and the red dashed line shows the data for the $N_{\rm p}=10^8$ simulation. The top left panel shows the disc surface density $\Sigma$. The top right panel shows the disc resolution $\left<h\right>/H$. The middle left panel shows the disc tilt profile $\beta$ scaled by the initial disc tilt $\beta_0$. The middle right panel shows the disc twist angle $\gamma$ in radians. The bottom left panel shows the warp amplitude $\psi = R\left|\partial\mathbi{l}/\partial R\right|$, where $\mathbi{l}$ is the unit vector pointing normal to the local orbital plane. The bottom right panel shows the magnitude of the numerical viscosity $\alpha_{\rm SS}^{\rm AV}$; the physical viscosity is zero in these simulations and thus this also represents the total viscosity present. The blue dotted line in the bottom right panel marks the line where $\alpha=H/R$. From these plots we can see that the simulations with $10^7$ and $10^8$ particles show similar behaviour, and that at $10^8$ particles the disc is in the wave-like regime ($\alpha<H/R$) at the time at which the disc breaks.}
	\label{FigA}
\end{figure}

We have followed the tearing simulations for sufficiently long timescales to briefly describe the time evolution. Focussing on the $N_{\rm p}=10^8$ case, we find that the innermost ring undergoes repeated precession around the black hole spin axis. Once the innermost ring has broken free, its radial extent is $1.1R_{\rm in} \lesssim R \lesssim 1.6R_{\rm in}$ and its inclination to the black hole spin is $\sim 20^\circ$. The ring undergoes $\sim 3$ full precession periods ($t_{\rm p} = 2\pi/\Omega_{\rm p}$) during which its azimuthal angle varies as $\gamma \approx \Omega_{\rm p}t$, where $\Omega_{\rm p}$ is the local (to the ring) Lense-Thirring precession frequency. During this time the ring's inclination remains approximately constant, decreasing by at most $\approx 10$ per cent, meaning that if the simulation were run for longer the ring would continue to precess many times before aligning to the black hole spin or being directly accreted. Across the ring the phase angle, $\gamma$, is essentially constant with radius, while the tilt angle is slightly decreasing with radius (from $\approx 23^\circ$ to $\approx 17^\circ$). There is also (at times later than depicted in Fig.~\ref{FigA}) a severe drop in surface density to the next ring of the disc (corresponding to 2-3 orders of magnitude in surface density) meaning that there is essentially no torque acting on the ring from the outer disc, allowing the repeated precession. This is caused by the extremely low viscosity present in the simulation, meaning that the timescale to refill the gap between the rings is much longer than the local precession timescale. The low viscosity also allows efficient wave communication and reflection (off the sharp transitions in surface density inside and outside of the ring) so that the ring can continue to precess.

\subsection{Long-term evolution of the tilt profile}
\label{longterm}

We would like to study the long-term evolution of the inner disc structure in the low-viscosity case to see whether the radial oscillations of the disc tilt are time steady, or whether they display secular behaviour such as alignment on longer timescales. In the linearised equations, the solutions that exhibit the radial tilt oscillations are steady with the torque arising from pressure balancing the precession of disc orbits\footnote{There is no local evolution of the disc angular momentum vector, with the external torque imposed on the disc carried to large radius. Thus there is a subtle distinction between discs of finite and infinite extent.}. However, in a fluid disc that accounts for additional effects, and in particular effects that act perpendicular to the torque arising from the pressure, it is not clear that such a solution can persist indefinitely. For example, typically the internal disc torques are comprised of components in each direction ($\mathbi{l}$, $\partial\mathbi{l}/\partial R$, and $\mathbi{l}\times\partial\mathbi{l}/\partial R$).

\begin{figure}
	\includegraphics[width=\columnwidth]{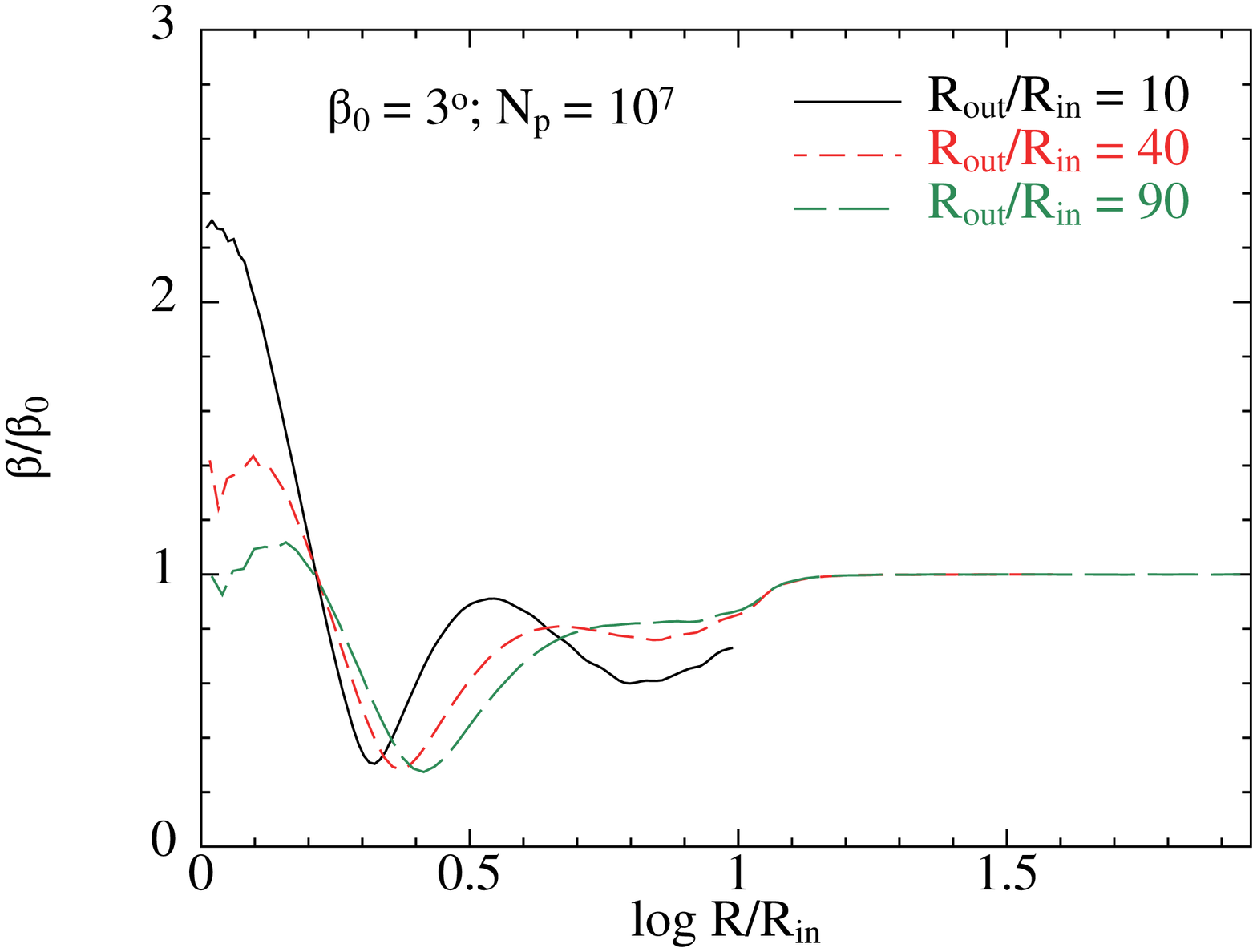}
	\includegraphics[width=\columnwidth]{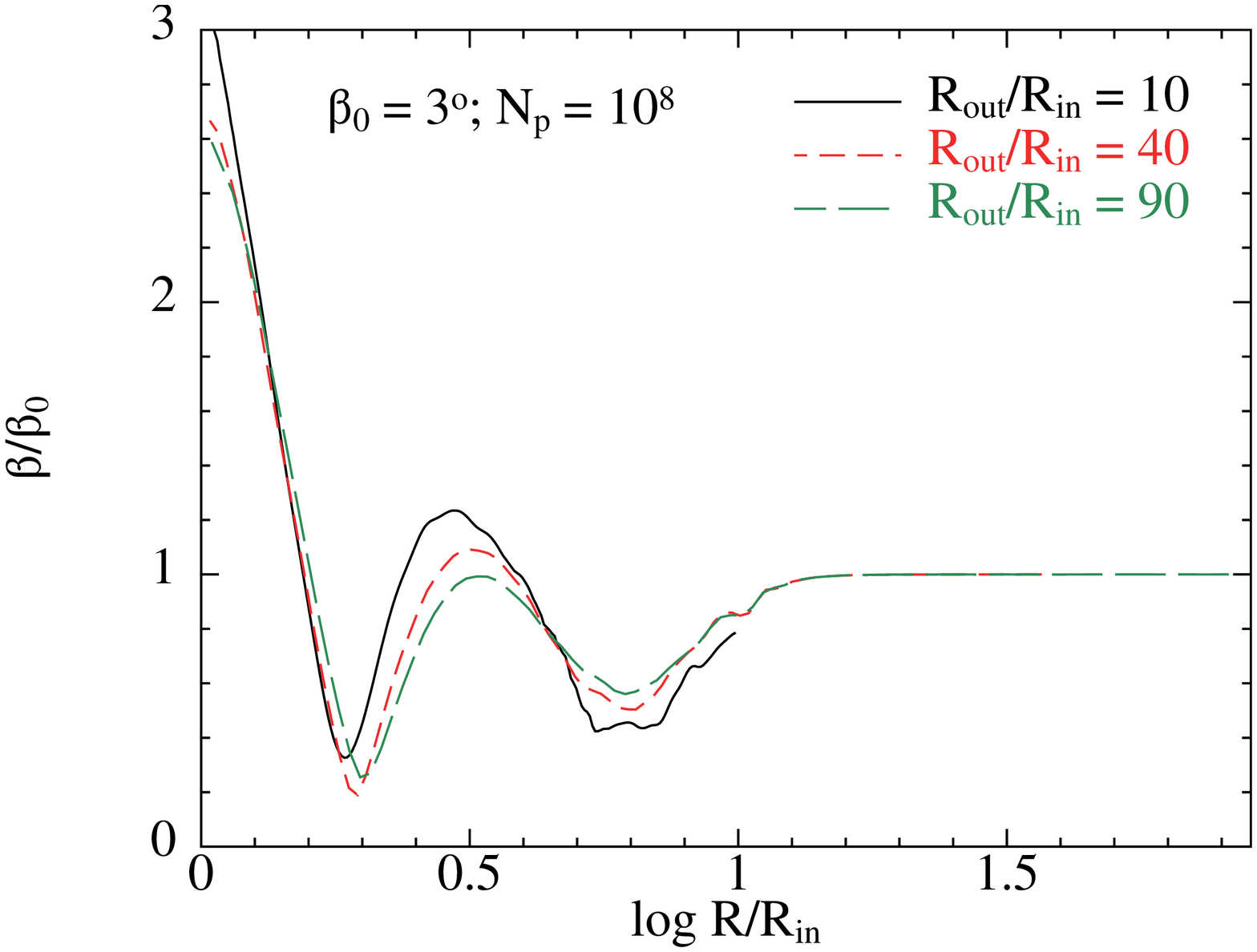}
	\caption{Tilt profiles at the end of the $\beta_0=3^\circ$ simulations, corresponding to $t=10^4\,GM/c^3$, with varying values of the outer disc radius $R_{\rm out}$. The top panel shows the simulations with $N_{\rm p}=10^7$ and the bottom panel shows the simulations with $N_{\rm p}=10^8$. The black solid lines show the data for the simulation with $R_{\rm out}/R_{\rm in}=10$, the red dashed line shows the simulation with $R_{\rm out}/R_{\rm in}=40$ and the green long-dashed line shows the simulation with $R_{\rm out}/R_{\rm in}=90$. Increasing the outer radius with a fixed particle number implies that the effective resolution decreases and thus the numerical viscosity increases (as $\left<h\right>/H$ increases). This is responsible for reducing the tilt in the inner disc in the upper panel as the outer radius is increased. In the lower panel we see that the results are similar when the resolution (number of particles) is higher.}
	\label{FigX}
\end{figure}

We can see from the results presented in, for example, Fig.~\ref{Fig1} that by the end of the simulation the tilt at the outer disc boundary has decayed significantly (to about 80\% of the original value). Therefore to explore the longer term behaviour we must extend the outer radius of the disc to ensure that the outer boundary is maintained over the duration of the simulation. However, modelling a larger radial range will lower the resolution of the simulation in the inner parts -- so it may not be possible with our setup to simulate faithfully the long-term evolution. We explore this here. In Fig.~\ref{FigX} we plot the tilt profiles of simulated discs at a time of $10^4\,GM/c^3$ with varying values of the outer radius. The black line corresponds to the simulations presented previously with $R_{\rm out}=40R_{\rm g}$ ($R_{\rm out}/R_{\rm in} = 10$), and then the red line corresponds to $R_{\rm out} = 160R_{\rm g}$ and the green line to $R_{\rm out}=360R_{\rm g}$. The top panel shows the simulations with $N_{\rm p}=10^7$ and the bottom panel shows the simulations with $N_{\rm p}=10^8$. From this figure we can see that at $10^7$ particles, the choice of outer boundary has a marked effect on the inner disc structure. This effect is primarily driven by the decreasing resolution, and hence increasing numerical viscosity, as the outer boundary is increased; this is evident from the decreased amplitude of the tilt at the inner disc edge. When the resolution is increased, with $N_{\rm p} = 10^8$, we see a more encouraging picture, with the disc structure essentially independent of the choice of outer boundary (Fig.~\ref{FigX}, bottom panel). Therefore, it should be possible to explore the long-term behaviour of the inner disc regions with such high resolution simulations. But, unfortunately, for reasons of computational cost, we are unable to continue these simulations to later times at present. We will return to this in the future.

\begin{figure}
	\includegraphics[width=\columnwidth]{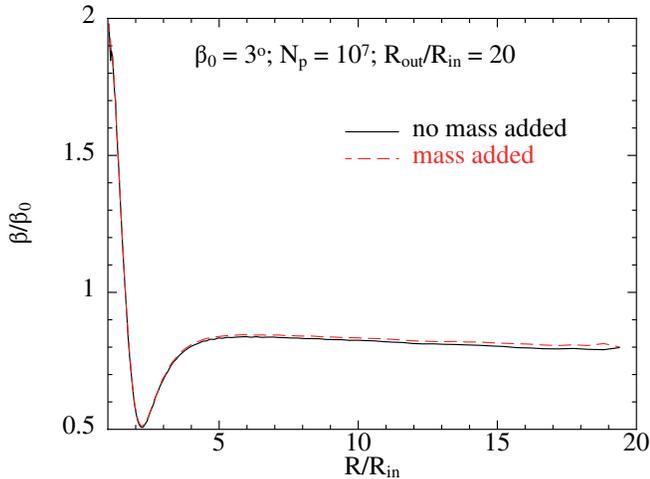}
	\caption{Tilt profiles at $t=10^4\,GM/c^3$ for the simulations exploring the addition of mass to discs with reduced radial range. Here we have $R_{\rm out}/R_{\rm in}=20$, $\beta_0=3^\circ$ and the initial discs are composed of $10^7$ particles. The initial disc surface density profile is determined by application of non-zero torque boundary conditions at both $R_{\rm in}$ and $R_{\rm out}$ (equation~\ref{steady}). The black line shows the tilt profile for the simulation where no mass was added to the disc, and the red dashed line shows the tilt profile for the simulation where mass was added to the outer disc regions at the rate required to keep the surface density profile steady on long timescales. There is little difference between the two models, indicating that the evolution of the outer disc tilt is primarily influenced by warp waves propagating from the inner regions. To model the long-term evolution of the disc inner regions it is therefore necessary to employ $R_{\rm out}/R_{\rm in} \gg 20$.}
	\label{FigY}
\end{figure}

An alternative, and lower computation cost, route to exploring the long-term behaviour is to model a restricted radial range for the disc with mass added to the outer disc regions over time. We have performed simulations to try this alternative. We detail in the Appendix the methodology we employ to inject mass (particles) into our disc over time. We set up two discs with $\beta_0 = 3^\circ$, $N_{\rm p} = 10^7$ and an outer disc radius of $R_{\rm out} = 80R_{\rm g}$. In these simulations we employ a physical viscosity of $\alpha_{\rm SS} = 0.02$ (using a direct Navier-Stokes viscosity; \citealt{Lodato:2010aa}) so that we may compute an expected accretion rate to determine the rate at which to add particles to the disc, and we also setup the disc surface density profile using equation~\ref{steady} (see Section~3 of \citealt{Nixon:2021aa}) to ensure that the disc surface density does not evolve significantly over time due to the inner or outer boundary conditions (for these simulations we apply a zero-torque boundary condition at both $R_{\rm in}$ and $R_{\rm out}$). In the first simulation we do not inject any mass with time as a control, and in the second simulation we inject mass at the required rate to support the disc surface density profile and we inject this mass into the original disc plane. We plot the tilt profiles for these two simulations in Fig.~\ref{FigY}. Unfortunately we again see that the outer disc tilt has evolved, even with the addition of mass to the outer disc regions. This demonstrates that in this instance the outer disc boundary should be set to a large value to correctly capture the long-term dynamics of the inner disc regions. This is necessary for the problem we are exploring here as the propagating warp waves are not damped by the time they propagate to radii of order $R_{\rm out}$ and thus they can affect the disc structure there. However, we expect that the methodology we have outlined here (in the Appendix) for adding mass to accretion disc simulations will be useful for simulating a variety of other astrophysical discs, including simulations in the diffusive warp propagation regime (where the propagating warp waves are damped and prevented from reaching the outer boundary) and simulations of circumbinary discs (e.g.\@ as suggested by \citealt{Heath:2020aa}).

\pagebreak

\section{Discussion}
\label{discussion}
\subsection{Simulation results}
We have presented numerical simulations of the evolution of low-viscosity warped discs around a spinning black hole. We have varied the tilt angle between the initial disc plane and the black hole spin, finding that for small angles ($\lesssim H/R$ radians) the disc evolves similarly to the solutions of the linearised equations for a wavelike warped disc (Fig.~\ref{Fig0}). At larger tilt angles we find that the results diverge from those at small tilt angles; for example at $\beta_0 = 10^\circ$ we find that the results are consistent with there being an enhanced local dissipation rate in regions of large warp amplitude. At $\beta_0 = 30^\circ$ we find that the disc breaks into discrete rings indicating that the internal communication of the warp was not sufficient to restrict the precession of disc orbits. The critical tilt values at which these effects occur depends on the disc angular semi-thickness and the black hole spin value.

We have shown that for low-viscosity discs around black holes with modest initial tilt angles, $\beta_0 \lesssim H/R$ radians, the steady radial oscillations of the disc tilt predicted by \cite{Ivanov:1997aa} and found in time-dependent 1D calculations by \cite{Lubow:2002aa} can be formed and sustained in 3D hydrodynamical simulations of the disc evolution. We find that high resolution $\left<h\right>/H \lesssim 0.1$ can be required to adequately describe the dynamics in the low-viscosity regime.

At high resolution, and thus correspondingly low numerical viscosity, the results of SPH simulations provide an adequate description of the propagation of warps in the linear regime. This is evident from the agreement between the solutions of the 1D linearised equations for a wavelike warp and the SPH solutions shown in Fig.~\ref{Fig0}. In this comparison the solutions are most discrepant in the innermost disc regions ($R/R_{\rm in}\lesssim 2$) which is where the numerical viscosity in the SPH simulations is largest, but it is also where the warp amplitude is largest. Specifically in this region $\psi \approx 0.1-0.2$ indicating that nonlinear effects may be important for initial tilt angles as small as $1^\circ$.

As the initial tilt angle is increased, and thus the warp amplitude through the disc increases, the simulations exhibit solutions that are consistent with those found at lower tilt angles but with increased local viscosity. This indicates that the disc is subject to enhanced local dissipation in regions of strong warping, and this acts to remove energy from the shearing motions in the warp and thus to reduce the warp amplitude. That the outcome of nonlinear dissipation may manifest as an increase in the local viscosity has been suggested by several authors, with the increased dissipation arising from shear instabilities \citep{Kumar:1993aa,Ivanov:1997aa} or from the parametric instability \citep{Gammie:2000aa,Bate:2000aa,Ogilvie:2013ab,Deng:2021aa}. As noted by \cite{Ivanov:1997aa} an increase in the local viscosity can eliminate some or all of the radial oscillatory tilt profile, and our simulations support this. The exact physical nature of the increased dissipation remains uncertain, but we have seen that it can result from only varying the initial tilt between simulations and thus the main physical difference is the strength of the shear within the warp which is indicated by the increase in the warp amplitude. Additional simulations and analysis beyond what we have presented here are required to determine exactly how the additional energy is extracted from the warp. For example, it may be that a simple model in which $\alpha = \alpha(\psi)$ is capable of bringing the linearised equations of motion into agreement with the nonlinear hydrodynamical simulations for tilt values that generate large values of $\psi$.

We have shown that discs with $\alpha < H/R$ can be unstable to the disc breaking instability \citep{Dogan:2018aa,Dogan:2020aa} leading to disc tearing \citep{Nixon:2012ad}. We observe repeated precession of the rings torn from the disc. Such repeated precession may be able to produce quasi-periodic oscillations (QPOs) in lightcurves; QPOs from Lense-Thirring precession was suggested by \cite{Stella:1998aa}, and has previously been discussed in the context of disc tearing by \cite{Nixon:2014aa} and \cite{Raj:2021ab}. We have confirmed in our simulations that, at the point at which the disc breaks, the viscosity is smaller than the disc angular semi-thickness and is thus in the classical wavelike warp propagation regime with $\alpha < H/R$. There have been recent claims that the observed disc structures in the proto-stellar/planetary system GW Ori are generated by disc tearing \citep{Kraus:2020aa} and such discs are generally thought to be in the wavelike warp propagation regime. Our findings support this possibility.

We have presented the methodology for adding mass over time to SPH simulations of discs to allow longer-term evolution of the discs to be explored (Appendix~\ref{inject}). This will be useful for modelling discs in several areas of astrophysics, including warped discs in the diffusive regime of warp propagation and circumbinary discs.

We also note that when the disc is retrograde with respect to the spin of the black hole the sign of the nodal precession is reversed while the sign of the apsidal precession is unchanged, and in this case the disc tilt no longer displays any oscillations with radius \citep{Ivanov:1997aa}. An example solution to the time-dependent 1D equations for this case is shown in Fig.~6 of \cite{Lubow:2002aa}. We have performed an additional simulation with $\beta_0 = 177^\circ$ (not depicted), and we find similar results; the tilt profile is smooth and counteraligned ($\beta_0 \approx 180^\circ$) at the inner boundary $R_{\rm in}$. 

\subsection{Observational implications}
We briefly discuss the observational relevance of our results. Our simulations suggest several possibilities for how a warp may provide a direct impact on the observable properties resulting from accretion discs around black holes \citep[see also, e.g.,][]{Nixon:2014aa,Raj:2021ab}. We have found that the inner disc may be significantly more tilted than the outer disc, and this may result in orientation-dependent increase or decrease in the flux reaching the observer from the inner disc regions. We have also seen that the tilt angle of the innermost region of the disc, $R < 2R_{\rm in}$, can oscillate with time as the disc settles into a steady state. Such time oscillations persist for longer in lower viscosity discs, and may lead to quasi-periodic features in the emission from the inner disc. Disc tearing in the low-viscosity regime results in rings that are able to precess several times before aligning to the black hole spin, providing an additional source of variability \citep[see][for a detailed discussion]{Raj:2021aa}.

It is worth highlighting the difference between discs with a finite outer boundary and discs in which the outer boundary tends to infinity \citep[cf.\@][]{King:2005aa}. In the latter case a steady warped solution for the (low-viscosity) disc shape is possible \citep{Ivanov:1997aa,Lubow:2002aa}. However, in the finite disc case, the torque applied to the disc (by the Lense-Thirring effect) is carried by waves to the outer boundary and may be reflected there. In this case the whole disc may be able to precess on a timescale given by the ratio of the disc angular momentum to the torque applied on the disc \citep[e.g.\@ equation 9 of ][]{Larwood:1996aa}. If the level of dissipation in the disc is non-zero, this leads to global alignment of the disc with the black hole spin vector over time \citep{Bate:2000aa,King:2013ab}. In low-viscosity discs with modest warp amplitude the alignment timescale can be longer than the precession timescale, and thus it may be possible for any jets emanating from the inner disc to precess on long timescales (given by the precession timescale of the whole disc). Observations of jet precessions \citep[e.g.\@][]{Aalto:2016aa,Aalto:2020aa} may indicate that in these systems the black hole accretion proceeds in the low-viscosity regime, as repeated precession of the inner disc regions is not routinely expected in the diffusive regime \citep{Nixon:2013aa}. This may indicate that $\alpha$ can be small, or that the disc is thick in these cases.

The steady radial oscillatory profile of the disc tilt, and particularly that $\beta(R_{\rm ISCO}) > 0$, may provide a plausible means of illuminating the outer disc with the central (e.g.\@ X-ray emitting) regions. The non-planar disc structure results in non-uniform illumination of the outer disc regions and this may contribute to the nature of the time lags observed in AGN reverberation mapping campaigns as suggested by \cite{Starkey:2017aa} and \cite{Fausnaugh:2018aa}.

Finally we note that if the disc shapes and properties that we have explored here can be linked directly with the observational properties of accreting black holes, then that will tell us important fundamental information about the accretion process in general. It is generally expected that black hole accretion discs are sufficiently thin and viscous that warps would be expected to propagate in a diffusive manner. For example, \cite{Martin:2019aa} summarise the observational evidence from different accreting systems in which the disc is expected to be fully ionised and they conclude that the evidence is consistent with $\alpha \sim 0.2-0.3$. However, there are notable exceptions in which one might expect the disc to have $\alpha < H/R$. These include the quiescent states of black hole X-ray binaries where the disc is not expected to be (fully) MRI active. In this case $\alpha$ is expected to be significantly smaller, and measurements from the quiescent states of Dwarf Novae (white dwarfs accreting mass from donor stars) suggest values of $\alpha \sim 0.01-0.001$ \citep[e.g.\@][see also \citealt{King:2013ab}]{Cannizzo:2012aa}. Another possibility is discs that are accreting near or above the Eddington limit; in this case the disc angular semi-thickness can be large, and approach unity in the inner disc regions. We therefore might expect a radial oscillatory tilt profile to be present in these cases. However, if these tilt profiles or other features of low-viscosity warped discs can be detected routinely in accreting black hole systems, that would place interesting constraints on accretion disc physics and particularly on the magnitude of $\alpha$ in these systems. We will therefore return to the observable consequences of these discs in the future to provide quantitative predictions on how these structures affect predictions of the disc emission.

\section{Conclusions}
\label{conclusions}
We have presented the first high-resolution and low-viscosity 3D hydrodynamical simulations of warped discs around spinning black holes that have successfully reproduced the main features of the analytical solutions to the 1D linearised equations presented by \cite{Ivanov:1997aa}, \cite{Demianski:1997aa} and \cite{Lubow:2002aa}. Previous attempts at this problem were hampered by insufficient spatial resolution in the numerical simulations, resulting in numerical viscosities that were too large to accurately model the warp wave propagation. We have presented and discussed several aspects of the simulations we have performed during this investigation. Our main conclusions are as follows:
\begin{itemize}
\item The steady radial oscillations in the disc tilt predicted by \cite{Ivanov:1997aa} can be established for long timescales in 3D hydrodynamical simulations of warped discs around spinning black holes when the disc tilt is sufficiently small.
\item By accounting for the zero-torque inner boundary condition we have shown that solutions to the 1D linearised fluid equations, and 3D hydrodynamical simulations, yield a tilt value at the disc inner edge that can be $\gtrsim 3\times$ the initial disc tilt.
\item As the initial disc tilt, and thus the warp amplitude, is increased between successive simulations the hydrodynamical solutions display additional dissipation that arises due to a transition from linear to nonlinear propagation of the warp. For example, by an inclination of $10^\circ$ (equivalent to several times the disc angular semi-thickness) the solution at high resolution is similar to the case where the inclination is $1^\circ$ but with the viscosity increased by approximately an order of magnitude.
\item We have presented a numerical simulation that exhibits disc breaking in which the magnitude of the numerical viscosity is measured to be small enough that $\alpha < H/R$ at the point at which the disc breaks. This provides the first clear evidence that disc breaking \citep{Nixon:2012ad,Dogan:2018aa} can also occur in the wavelike regime.
\item We have shown that to adequately establish the long term evolution of the warp structure of the inner disc regions it is necessary to employ an outer disc boundary at sufficiently large radius that propagating waves do not have time to reflect off the outer boundary and return to the central regions of the disc. Determining the long term behaviour will require additional simulations to those presented here.
\item Finally, we have presented and used methodology for the injection of mass into SPH simulations of discs that will be useful for accretion discs in a variety of astrophysical contexts.
\end{itemize}

\section{Acknowledgments}
We thank the referee for a positive and useful report. We thank Jim Pringle for useful discussions and detailed comments on the manuscript. We thank Gordon Ogilvie for detailed comments on the manuscript. CJN acknowledges funding from the European Union’s Horizon 2020 research and innovation program under the Marie Sk\l{}odowska-Curie grant agreement No 823823 (Dustbusters RISE project). This research used the ALICE High Performance Computing Facility at the University of Leicester. This work was performed using the DiRAC Data Intensive service at Leicester, operated by the University of Leicester IT Services, which forms part of the STFC DiRAC HPC Facility (\url{www.dirac.ac.uk}). The equipment was funded by BEIS capital funding via STFC capital grants ST/K000373/1 and ST/R002363/1 and STFC DiRAC Operations grant ST/R001014/1. DiRAC is part of the National e-Infrastructure. We used {\sc splash} \citep{Price:2007aa} for the figures.

\bibliographystyle{aasjournal}
\bibliography{nixon}

\appendix
\section{Mass injection for SPH discs}
\label{inject}

In this Appendix we detail the method we used for adding mass to the discs over time in Section~\ref{longterm}. For the initial conditions of these simulations we use a disc $\Sigma(R)$ that is time-steady in a 1D model and straightforward to implement. To do this we impose boundary conditions at both $R_{\rm in}$ and $R_{\rm out}$. We take both boundary conditions to be zero-torque, i.e.\@ that any mass reaching $R<R_{\rm in}$ or $R>R_{\rm out}$ is removed along with its angular momentum. To keep the disc steady we must add mass at a rate ${\dot M}$ at a radius $R_{\rm add}$. This gives (Section 3, \citealt{Nixon:2021aa})
\begin{equation}\label{steady}
\Sigma(R) =
\begin{dcases}
  \frac{{\dot M}_{\rm add}}{3\pi\nu(R)}\left[1-\left(\frac{R_{\rm in}}{R}\right)^{1/2}\right]\frac{R_{\rm out}^{1/2}-R_{\rm add}^{1/2}}{R_{\rm out}^{1/2}-R_{\rm in}^{1/2}}~~{\rm for}~~R\le R_{\rm add}\\
  \frac{{\dot M}_{\rm add}}{3\pi\nu(R)}\left[\left(\frac{R_{\rm out}}{R}\right)^{1/2}-1\right]\frac{R_{\rm add}^{1/2}-R_{\rm in}^{1/2}}{R_{\rm out}^{1/2}-R_{\rm in}^{1/2}}~~{\rm for}~~R>R_{\rm add}
\end{dcases}
\end{equation}
Note that we have assumed here that mass is added at $R_{\rm add}$ with zero width. In the SPH simulations we choose to smooth the addition of mass over a finite, non-zero radial width (see below) This affects the steady surface density profile as we will see below. Similarly, in 1D calculations, in which $V_R \ll V_\phi$ (and thus circular orbits are assumed), the boundary conditions can be enforced exactly. However, in a fluid disc particles arrive at the boundary with small, but nonzero, eccentricity. This means that the angular momentum of the particles when they are removed does not exactly correspond to the angular momentum of a circular orbit at the boundary. In the high resolution limit, we would therefore expect particles arriving at $R_{\rm in}$ ($R_{\rm out}$) to have a slight excess (deficit) of angular momentum with respect to circular orbits at the boundary. However, for low enough resolution a particle approaching the inner boundary may be subject to sufficient numerical viscosity---exacerbated by the surface density approaching zero at the inner boundary---that enough angular momentum is extracted from the particle's orbit that upon arriving at $R_{\rm in}$ the particle has $L < L_{\rm in,circ}$. In this case the particle has given up to the disc some of the angular momentum that should have been accreted (in the notation of \citealt{Nixon:2021aa} this corresponds to $0 < f \ll 1$). The same process occurs, albeit with the sign of the angular momentum exchange reversed, at the outer boundary. For $H/R \ll 1$, and thus $V_R \ll V_\phi$, we can therefore expect that at low resolution the surface density at the boundaries will be non-zero, but that as the resolution is increased the surface density there will move closer to zero. When $V_R$ at the boundary is larger, for example when the disc angular semi-thickness is increased, we find that the surface density at the boundaries is also increased. The corollary of this is that when comparing 1D calculations to 3D hydrodynamical simulations we can expect differences in the surface density profiles, particularly at and near the boundaries. Some of these differences are numerical (e.g.\@ the increasing numerical viscosity near the boundaries in the 3D simulations) and some are physical (e.g.\@ the inclusion of pressure gradients which alter the rotation profile from exactly Keplerian and cause non-negligible particle eccentricities near the boundaries of the 3D simulations).

We setup the initial disc by placing particles using the usual Monte-Carlo method to achieve the desired density profile. Then over time we add particles to the disc. To do this we must provide at each timestep the number of particles to inject into the simulation and each particle's position, velocity, and estimated smoothing length. For the number of particles we calculate the mass added to the disc over the timestep, and save the left over fraction of a single particle's mass and add this to the mass calculated in the next timestep. For the estimate of the smoothing length we could take an average over the local particles, but we deem this too computationally expensive to search for nearby particles. Similarly, we could use the relation between the density and smoothing length $h = \eta (m_{\rm p}/\rho)^{1/3}$ (where $\eta$ specifies the smoothing length in units of the mean (local) particle spacing and $m_{\rm p}$ is the mass of a single particle) combined with $\rho \sim \Sigma/H$ to yield $h_{\rm est} = \eta(Hm_{\rm p}/\Sigma)^{1/3}$, but this requires knowledge of the local $\Sigma$ which may vary with time. So instead we take the simple approach of using $h_{\rm est} = H$, and note that the Newton-Raphson iteration (see, e.g., \citealt{Price:2012aa}) between $\rho$ and $h$---which is performed every timestep---quickly yields the correct value of $h$ with little additional computational cost for the small number of particles injected per timestep. We take a similar approach for the velocity of the injected particles. It is possible to compute the local radial pressure gradient, and use this to determine the correct rotation velocity with which to add the particles. But instead, for simplicity, we add the particles on circular orbits with the local Keplerian azimuthal velocity. Finally we place the particles at a position determined by (1) a vertical height using the same Monte-Carlo method for the Gaussian vertical distribution in the initial disc, (2) an azimuthal angle drawn from a random number generator between zero and $2\pi$, and (3) a radius using a Monte-Carlo method to distribute the particles following a cosine-bell (Hahn function) centred on $R_{\rm add}$ and extending from $R_{\rm add}-\Delta R$ to $R_{\rm add}+\Delta R$, with the width $\Delta R = w_{\rm add}H(R_{\rm add})$. We take $w_{\rm add} = 3$ for the simulations we perform here.

\begin{figure*}
	\centering
	\includegraphics[width=0.4\columnwidth]{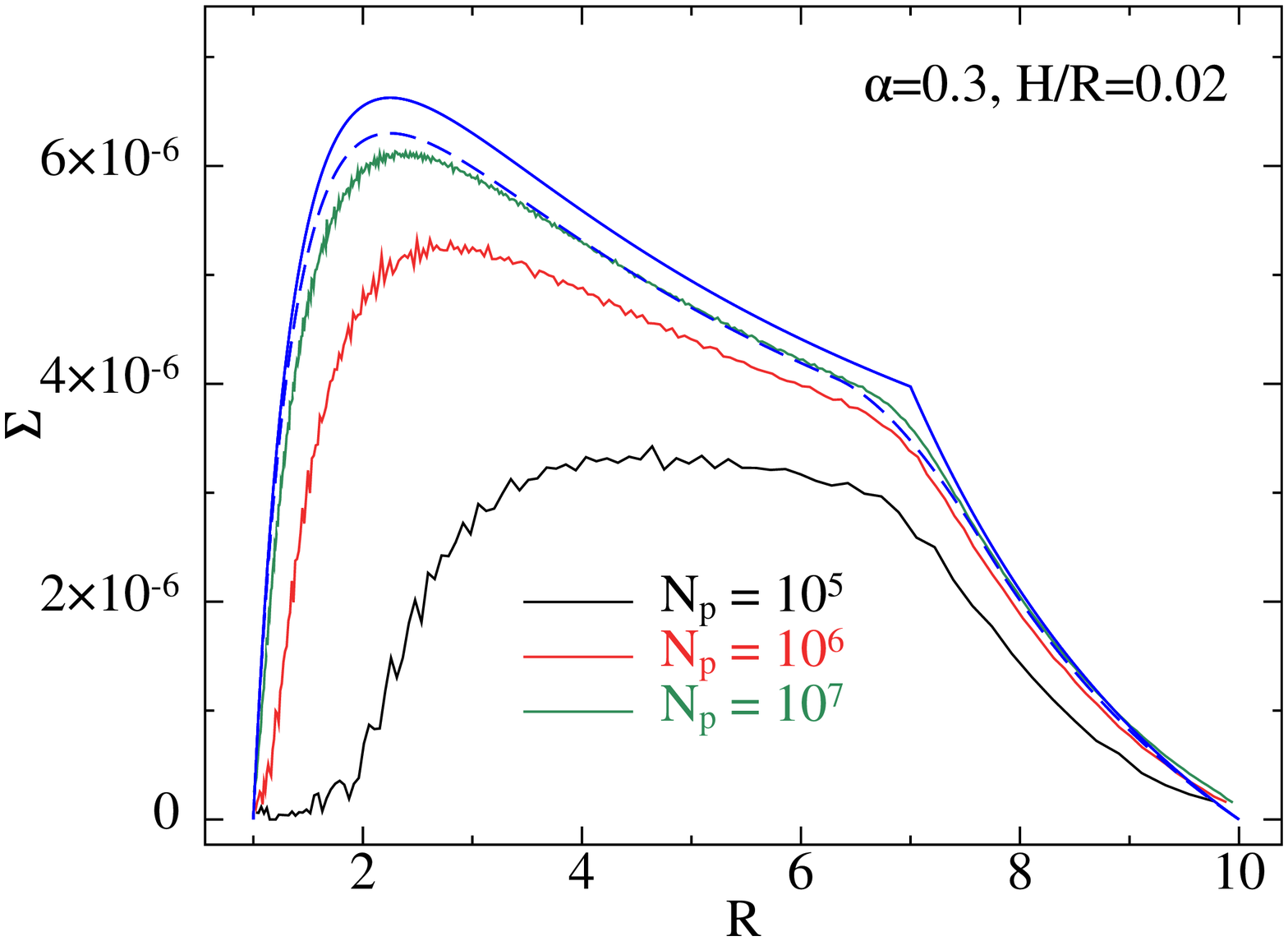}\hspace{0.5in}
	\includegraphics[width=0.4\columnwidth]{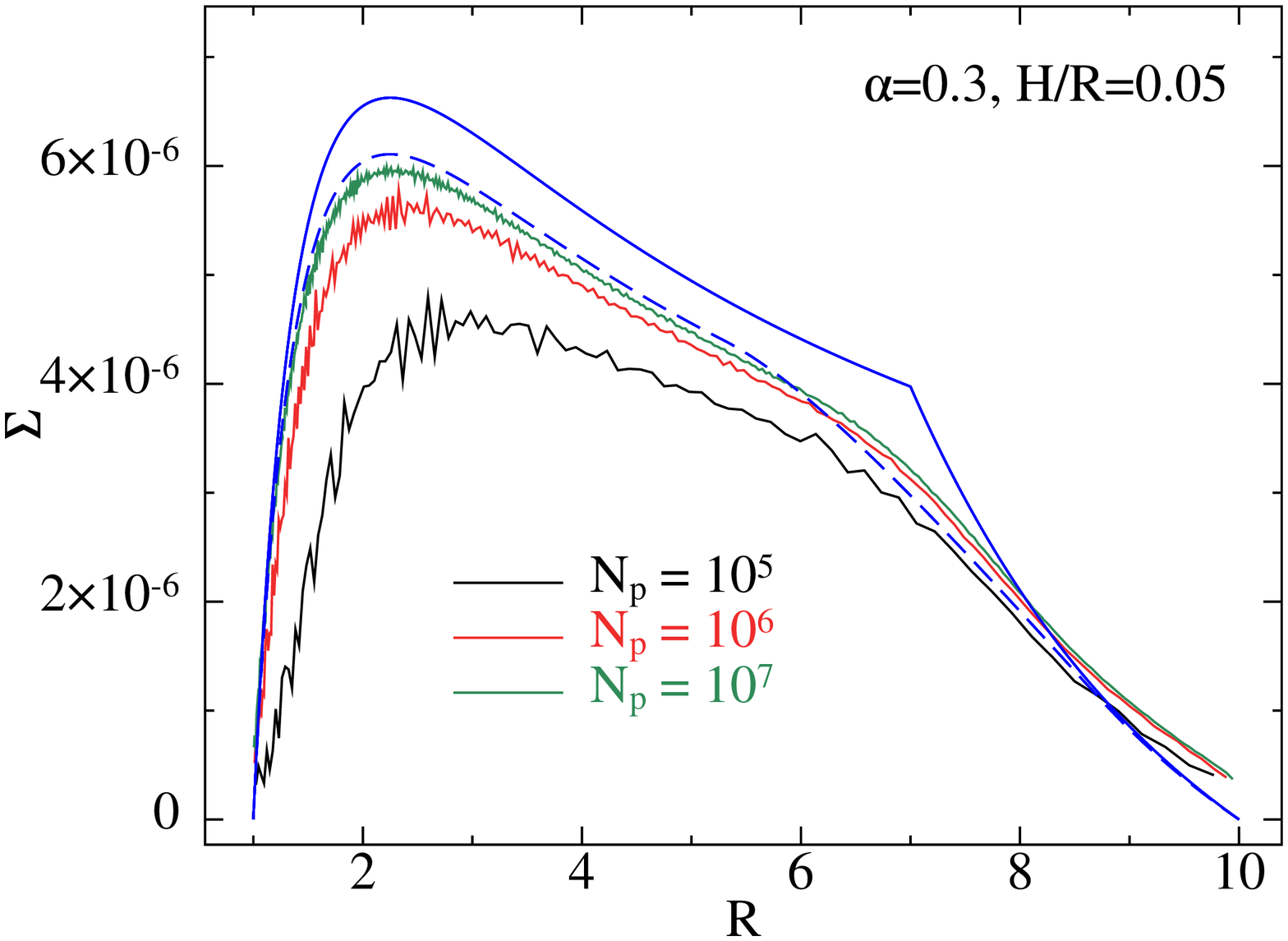}
	\caption{Surface density profiles of the discs with mass added over time. The left panel shows the discs with $H/R=0.02$ and the right panel shows the discs with $H/R=0.05$. The blue solid line shows the initial profile (equation~\ref{steady}), which assumes a $\delta$-function for the mass input (i.e.\@ zero width). The blue dashed line shows the result of integrating the 1D diffusion equation for a disc with the relevant parameters and including the cosine-bell smoothed mass input. The black, red and green lines correspond to the discs that start with $N_{\rm p} = 10^5$, $10^6$ and $10^7$ respectively, and are plotted at a time corresponding to $9M_{\rm d}/{\dot M}$. The surface density data for the SPH simulations are computed by placing the particles into logarithmically spaced radial bins, and then dividing the total mass by the area of the annulus. For $N_{\rm p}=10^5$ we use 100 radial bins, for $N_{\rm p}=10^6$ we use 200 radial bins, and for $N_{\rm p}=10^7$ we use 400 radial bins. As the resolution is increased there is closer agreement with the prediction from the 1D diffusion equation.}
	\label{FigApp}
\end{figure*}

To demonstrate the injection of particles over time into an SPH discs we simulate a planar, Keplerian disc with $R_{\rm in} = 1$, $R_{\rm out}=10$, $R_{\rm add} = 7$, a locally isothermal equation of state with sound speed power-law with $q=0.25$, an initial disc mass of $M_{\rm d} = 0.001$ in units of the central object's mass, and a Shakura-Sunyaev viscosity with $\alpha=0.3$ imposed as a direct Navier-Stokes viscosity. We evolve the disc with mass added at a rate ${\dot M}$ in the manner described above for a time corresponding $9M_{\rm d}/{\dot M}$, such that the total number of particles involved in the simulation corresponds to $10\times$ the number in the initial disc. For these disc parameters, this timescale is sufficient for the disc to approach a steady state and corresponds to approximately a viscous timescale of the disc at $R_{\rm add}$; note that as we are injecting mass near the outer edge of the disc most of the mass flows off the outer boundary \citep[see equation 22 of][]{Nixon:2021aa}. We perform simulations with the initial disc composed of $10^5$, $10^6$ and $10^7$ particles. We perform two sets of these simulations, one with $H/R=0.02$ at $R_{\rm in}$ and one with $H/R=0.05$ at $R_{\rm in}$. The results of these simulations are shown in Fig.~\ref{FigApp}. We plot the initial surface density profile (equation~\ref{steady}; blue solid line), the prediction found by evolving the 1D diffusion equation for a disc with mass input in the form of the cosine-bell as has been applied to the SPH simulations (blue dashed line), and the (approximately) steady-state SPH simulations results with the initial disc composed of $10^5$ particles (black line), $10^6$ particles (red line) and $10^7$ particles (green line). The left panel shows $H/R = 0.02$ and the right panel shows $H/R=0.05$.

Fig.~\ref{FigApp} shows that at low resolution the disc surface density profile decays significantly before reaching a steady state. This is principally due to the large numerical viscosity in this case. As the resolution is increased the numerical viscosity becomes small, and insignificant when compared to the physical viscosity imposed in these simulations ($\alpha=0.3$). For lower values of the physical viscosity the ratio of numerical to physical viscosity will be increased and the results more discrepant from the target surface density profile, meaning that higher resolution would be required to achieve the same results in this case. In Fig.~\ref{FigApp} we can see that the surface density is not zero at the outer boundary as expected from the discussion above. This is also true at the inner boundary but to a much lesser extent, with the surface density close to zero there. We can see that the disc surface density is above the predicted value at and just outside the injection region. This is perhaps signifying that the addition of matter at the Keplerian rotation velocity is not sufficient here. This is because adding material with too much angular momentum compared to the local flow results in a torque that expels material outwards. It is likely that accounting for the disc rotation profile (which is sub-Keplerian due to the radial pressure gradient) would improve the fit here, and we will explore this in future work. We note that it may also be possible to achieve a closer agreement between the expected and simulated surface density profiles, particularly at low resolution, by accounting for the impact of the numerical viscosity on the mass flow rates through the inner and outer boundary. A zeroth order correction would be to adjust the mass addition rate ${\dot M}$ to account for this.

\end{document}